\documentclass[aps,amsmath,twocolumn,amssymb,titlepage,10pt]{revtex4-1}
\usepackage[T1]{fontenc}
\usepackage[utf8]{inputenc}
\usepackage{amsmath}
\usepackage{braket}
\usepackage{amsfonts}
\usepackage{comment}
\usepackage{graphicx}
\usepackage{hyperref}
\usepackage[capitalize]{cleveref}
\usepackage{amssymb}
\usepackage{amsmath}
\usepackage{mathtools}
\usepackage{todonotes}
\DeclareMathOperator{\Tr}{Tr}
\newcommand{\ketbra}[2]{\ket{#1}\!\bra{#2}}
\newcommand{\ua}{\uparrow}
\newcommand{\da}{\downarrow}

\begin{document}
\title{Cluster quantum Monte Carlo 
study of two-dimensional\\ weakly-coupled frustrated trimer
antiferromagnets}
\author{Lukas Weber}
\author{Nils Caci}
\author{Stefan Wessel}
\affiliation{Institute for Theoretical Solid State Physics, RWTH Aachen University, JARA Fundamentals of Future Information Technology, and \\ JARA Center for Simulation and Data Science, 52056 Aachen, Germany}
\begin{abstract}
We report results from spin trimer-based cluster quantum Monte Carlo simulations for the thermodynamic properties of two-dimensional
frustrated quantum antiferromagnets that are composed of  weakly-coupled three-spin (trimer) clusters. In particular, we consider the  spin-1/2 kagome lattice with  a strong breathing distortion, and the triangle-square lattice model proposed previously  for the cuprate La${}_4$Cu${}_3$MoO${}_{12}$. For both cases, we demonstrate that an appropriately chosen trimer-based computational basis allows us to  significantly reduce  the quantum Monte Carlo sign problem down to the low-temperature regime. Besides exploring the thermodynamic behavior for the triangle-square lattice model we also assess a mean field theory-based prediction for the onset of chiral order. For the breathing distorted kagome lattice model, we observe a robust two-peak structure in the specific heat, both in the quantum spin liquid and the lattice-nematic regimes. 
\end{abstract}
\maketitle
\section{Introduction}
Triangular clusters form one of the most basic building blocks of frustrated magnets in two dimensions~\cite{Starykh2015}, including the paradigmatic Heisenberg antiferromagnets on the triangular or kagome lattice. In particular the $S=1/2$ Heisenberg antiferromagnet on the kagome lattice, which can be interpreted as a a lattice of corner-sharing triangles, has attracted considerable interest as a promising system for realizing a quantum spin liquid~\cite{Savary2016,Yan2011,Depenbrock2012,Jiang2012,He2017} in its groundstate. In most candidate materials, the magnetic exchange couplings do however not realize a perfect kagome lattice and are instead distorted~\cite{Norman2016}, which can have significant impact on the low temperature physics, e.g., by stabilizing magnetic order~\cite{Schnyder2008}, precluding the formation of a quantum spin liquid.

\begin{figure}
	\includegraphics{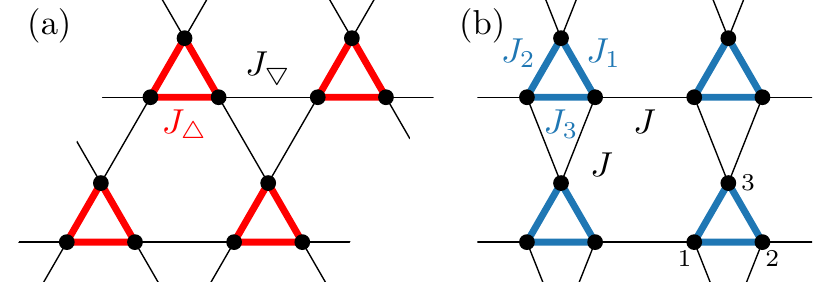}
	\caption{(a)
	The breathing kagome lattice with intratrimer bonds $J_\triangle$ along the upwards triangles and intertrimer bonds $J_\bigtriangledown$ along the downwards triangles.
	(b) 
	The triangle-square lattice with the intratrimer bonds $J_1$, $J_2$, and $J_3$ as well as the intertrimer bonds $J$. The $x$ ($y$) lattice direction points to the right (upwards).
	}
	\label{fig:lattices}
\end{figure}

This is, however, not true for all kinds of distortions, as seen for example in the “breathing-distorted” kagome lattice, where upward and downward triangles both remain equilateral but inequivalent to each other (\cref{fig:lattices}(a))~\cite{Mila98,Aidoudi2011,Honecker11}. Even under strong such breathing distortion, a quantum spin-liquid phase of the kagome lattice is expected to be stable~\cite{Repellin2017,Jahromi2020}. Such a distortion is indeed observed in the spin-1/2 vanadium oxyfluoride 
[NH${}_4$]${}_2$[C${}_7$H${}_{14}$N][V${}_7$O${}_6$F${}_{18}$],
which shows no sign of order or spin freezing down to low temperatures~\cite{Aidoudi2011,Clark2013,Orain2017}.
It is thus important to accurately examine the ground state and low-temperature thermodynamic behavior of such distorted model systems. 

Unbiased numerical studies of both the undistorted and the breathing distorted kagome lattice have so far been based mainly on exact diagonalization (ED) calculations~\cite{Yan2011,Sugiura13,Laeuchli19,Schnack18} on (relatively) small clusters, tensor-network methods such as the density matrix renormalization group (DMRG)~\cite{Yan11,Depenbrock2012,Repellin2017,He17}  or infinite projected-entangled pair states (iPEPS) and related approaches~\cite{Liao17,Chen18,Jahromi2020}, which do however suffer in the case of DMRG from finite-circumference cylinder geometries and in the case of iPEPS from limited accessible bond dimensions. Unbiased quantum Monte Carlo methods such as the stochastic series expansion (SSE)~\cite{Sandvik1991,Sandvik1992,Sandvik1999,Syljuasen2002,Alet2005} typically suffer from the negative sign problem~\cite{Henelius00, Troyer2005} in the face of frustration. Averting this basis-dependent sign problem is however possible for a set of special models by performing a local change of basis. A notable example for this is the class of highly-frustrated Heisenberg antiferromagnets on lattices with “fully frustrated” interactions, where changing to a suitable cluster basis removes the sign-problem~\cite{Nakamura1998,Alet2016,Honecker2016,Ng2017,Stapmanns2018,Weber2021}.

In this work, we show how a suitable cluster basis can also be beneficial in the case of weakly coupled trimer clusters. Here, the sign problem is in general still present but can be reduced to a degree where unbiased calculations of the thermodynamic quantities become possible down to the low-temperature regime. We will perform these calculations for the $S=1/2$ Heisenberg antiferromagnet on two basic lattices of coupled trimers, highlighting the strengths and shortcomings of this approach. First, we will consider a square lattice of triangle clusters (the triangle-square lattice, \cref{fig:lattices}(b)), where the triangular clusters are arranged on a bipartite parent lattice so that the frustration predominantly arises within the clusters themselves. This model Hamiltonian has been put forward~\cite{Wessel2001,Wang2001} to describe the frustrated quantum magnetism observed in the cuprate-based compound La${}_4$Cu${}_3$MoO${}_{12}$.
Second, we will then return to the kagome lattice with strong breathing distortion (\cref{fig:lattices}(a)) as a more challenging example where also the interaction between trimers takes place on a highly frustrated triangular parent lattice~\cite{Repellin2017,Jahromi2020}.

The remainder of this article is  structured as follows. In \cref{sec:basis} we outline the details of the different cluster bases we employ in our simulation. In \cref{sec:ts}, we present our results for the triangle-square lattice, which are contrasted in \cref{sec:bk} to those for the breathing-distorted kagome lattice. Finally, in \cref{sec:conc}, we draw a conclusion on our results and provide an outlook.

\section{Cluster bases}
\label{sec:basis}
In this paper, we employ and compare three different computational bases that correspond to the eigenbasis of different physical single-trimer operators. The first is the conventional single-spin $S_z$-basis~\cite{Sandvik1991}, in which  the $S^z$ component of each trimer spin is diagonal,
\begin{equation*}
	S^z_{\Delta,\alpha} \ket{m_1,m_2,m_3} = m_\alpha \ket{m_1,m_2,m_3},\quad \alpha = 1,2,3.
\end{equation*}
The second is the dimer basis, $\ket{l_{12}, m_{12}, m_3}$, which diagonalizes the total spin, $\mathbf{S}_{\Delta,12}^2 = (\mathbf{S}_{\Delta,1} + \mathbf{S}_{\Delta,2})^2$, and magnetization, $S_{\Delta,12}^z = S^z_{\Delta,1} + S^z_{\Delta,2}$, along one bond of the trimer (taken here to be the one connecting $\mathbf{S}_1$  and $\mathbf{S}_2$) and leaves the third spin untouched,
\begin{align}
	\ket{1, 1, m_3} &= \ket{\ua, \ua, m_3},~\ket{1, -1, m_3} = \ket{\da, \da, m_3},\nonumber\\
	\ket{1, 0, m_3} &= (\ket{\ua,\da} + \ket{\da,\ua}) \otimes \ket{m_3}/\sqrt{2}, \nonumber\\
	\ket{0, 0, m_3} &= (\ket{\ua,\da} - \ket{\da,\ua}) \otimes \ket{m_3}/\sqrt{2}.
\end{align}

Finally we consider the trimer basis,
where the total spin, $\mathbf{S}_\Delta^2 = (\sum_{\alpha=1}^3 \mathbf{S}_{\Delta,\alpha})^2$, and total magnetization, $S^z_\Delta = \sum_{\alpha=1}^3 S_{\Delta,\alpha}^z$ of the full trimer are diagonal. In addition to these two, a third operator is needed to completely distinguish the 8 states on the trimer. This operator is not uniquely determined, but can be readily constructed from symmetry considerations. First, choosing an SU(2) symmetric operator ensures that it has a common eigenbasis with $\mathbf{S}^2_\Delta$ and $S^z_\Delta$. Second, by additionally requiring time-reversal symmetry, its eigenbasis (and thus the trimer basis) can be chosen to be real-valued with respect to the single-spin $S^z$ basis. Therefore, the Hamiltonian, which is also real-valued in the $S^z$ basis, will not acquire a phase problem by transformation to the trimer basis \footnote{Even with a complex basis, one can, in principle, retain real-valued weights within the SSE and related algorithms due to pairs of configurations with cancelling imaginary parts \cite{Hen2021}. In this work, however, we concentrate on real-valued bases.}. The simplest way of fulfilling these condition is to use one of the dimer total spin operators, such as $(\mathbf{S}_1 + \mathbf{S}_2)^2$, which we do in the following. Therefore, for the trimer basis $\ket{l_\Delta, m_\Delta, l_{\Delta,12}}$, we have three quantum numbers and
\begin{align}
	\ket{3/2, +3/2, 1} &= \ket{\ua\ua\ua},\quad \ket{3/2, -3/2 ,1} = \ket{\da\da\da},\nonumber\\
	\ket{3/2, +1/2, 1} &=  (\ket{\ua\ua\da} + \ket{\ua\da\ua} + \ket{\da\ua\ua})/\sqrt{3},\nonumber\\
	\ket{3/2, -1/2, 1} &=  (\ket{\da\da\ua} + \ket{\da\ua\da} + \ket{\ua\da\da})/\sqrt{3},\nonumber\\
	\ket{1/2, +1/2, 0} &= (\ket{\ua\da} - \ket{\da\ua}) \otimes \ket{\uparrow}/\sqrt{2},\nonumber\\
	\ket{1/2, -1/2, 0} &= (\ket{\da\ua} - \ket{\ua\da}) \otimes \ket{\downarrow}/\sqrt{2},\nonumber\\
	\ket{1/2, +1/2, 1} &= (\ket{\ua\da\ua} + \ket{\da\ua\ua} - 2 \ket{\ua\ua\da})/\sqrt{6},\nonumber\\
	\ket{1/2, -1/2, 1} &= (\ket{\da\ua\da} + \ket{\ua\da\da} - 2 \ket{\da\da\ua})/\sqrt{6}.
\end{align}
For the Hamiltonian of uncoupled trimers, the single-spin basis has a sign problem. The dimer basis is sign-free under the condition that the decoupled trimers are mirror symmetric perpendicular to the dimer it singles out. By contrast, the trimer basis can be made sign-free in any case by choosing an appropriate third operator so that it coincides with the eigenbasis of a single trimer.

Once an intertrimer coupling is introduced, in general, all three bases are subject to the sign problem. The severity of this sign problem depends on the model studied and can also differ largely between different bases. For example, for the specific case of the fully-frustrated trilayer  (FFTL) model considered in Ref.~\cite{Weber2021}, the sign problem is completely eliminated only in the trimer basis (the FFTL is obtained upon adding in \cref{fig:lattices}(b) couplings of strength $J$ between all spins belonging to nearest-neighbor trimers). 
In the following, we will show how, in contrast to the single-spin basis, the dimer and trimer bases remain useful in the regime of weak intertrimer coupling and allow us to resolve the thermodynamics of the two trimer magnets that we examine here. 

\section{Triangle-square lattice}
In this section, we first consider the triangle-square lattice model of coupled spin trimers. In this system, the trimers are arranged on a biparite square lattice, so that the magnetic frustration resides  within the individial trimers.  The ground state phase diagram of this model has been analyzed based on ED  and mean-field theory~\cite{Wessel2001,Wang2001}. As detailed further in Sec.~\ref{Sec:TSGSO}, in the weak $J'$-regime, the ground state phase diagram is composed of various magnetically ordered regimes, depending on the relative strength of the three intratrimer couplings. In the following we consider finite-size systems with periodic boundary  conditions, where the number of spins relates via $N=3L^3$ to the linear system size $L$  of the square lattice. 
\label{sec:ts}
\subsection{Sign}
\begin{figure}
	\includegraphics{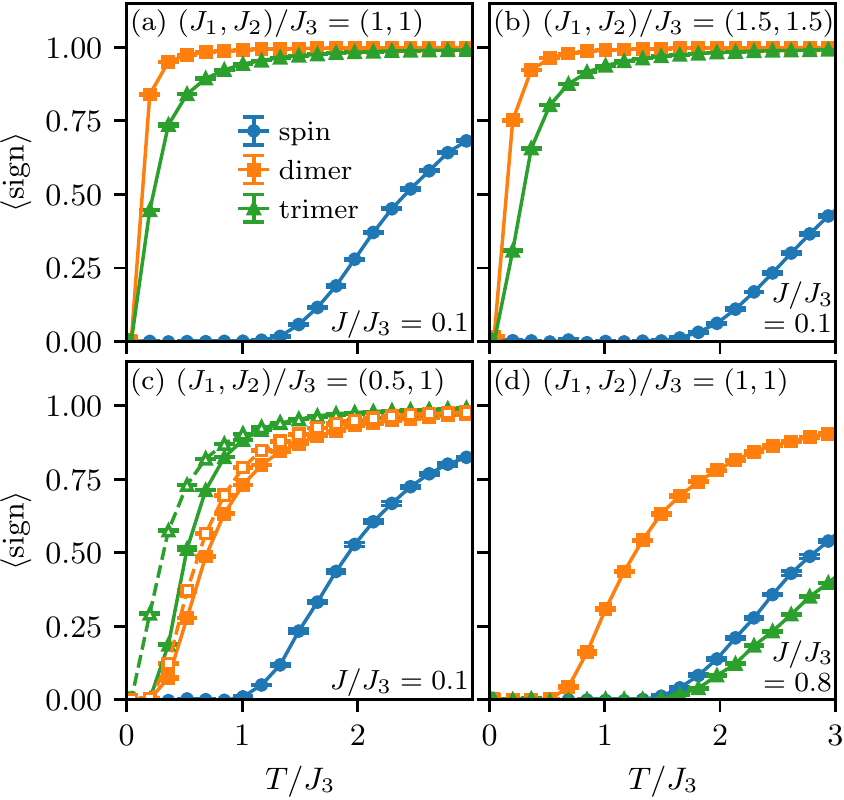}
	\caption{Average configuration sign, $\langle \mathrm{sign}\rangle$, of the $L=12$ triangle-square lattice in different computational bases for different values of the intratrimer couplings $J_{1/2/3}$ and the intertrimer coupling $J$. In panels (a--d) solid lines show the dimer and trimer basis using the total spin of the $\mathbf{S}_1$-$\mathbf{S}_2$ dimer as a quantum number. In panel (c), the dashed lines show rotated versions of these bases using the $\mathbf{S}_2$-$\mathbf{S}_3$ dimer, allowing for an improved sign in this case.}
	\label{fig:triangle_square_sign}
\end{figure}
We start our discussion of the triangle-square lattice by comparing the average configuration sign, 
$\langle \mathrm{sign}\rangle$~~\cite{Henelius00, Troyer2005, Honecker2016}, for the three different bases and at different points in the parameter space of the intradimer couplings.  (\cref{fig:triangle_square_sign}). Generally, the average sign decays to zero as temperature goes to zero, but the scale of this decay depends on the basis. For the single-spin basis it  is set by the dominant of the intratrimer couplings $J_{1}$, $J_{2}$,  $J_{3}$. For weakly coupled trimers (\cref{fig:triangle_square_sign}(a--c)), the dimer and trimer bases greatly outperform the single spin basis, retaining a robust sign down to $T\sim J$. Along the symmetric line $J_1 =J_2$, the dimer basis is slightly more favorable  than the trimer basis and off the symmetric line (\cref{fig:triangle_square_sign}(c)), the trimer basis is favorable. In the special case where $J_1\ne J_2$ but $J_2=J_3$ a rotated version of the dimer and trimer basis, using the $S_2$-$S_3$ instead of the $S_1$-$S_2$ dimer, can be used to further enhance the average sign (\cref{fig:triangle_square_sign}).
For stronger couplings (\cref{fig:triangle_square_sign}(d)), the trimer basis becomes less effective than the single spin basis while the dimer basis retains a robust sign down to $T\sim J$ (in practice, our SSE simulations remain feasible for $\langle \mathrm{sign}\rangle$  larger than  $O(10^{-2})$).

This behavior can be intuitively understood by considering the ways in which the triangle-square lattice can be extended to eliminate the sign-problem in the respective bases. For the single-spin basis, no such extension exists. The only way to remove the sign-problem is to remove couplings -- both inter- and intratrimer -- until there is no more geometric frustration. For the dimer basis, if $J_1 = J_2$, the model becomes sign-free upon adding further interactions of strength $J$ among all spins that belong to neighboring horizontal dimers so that 
that the model turns into rows of fully frustrated two-leg ladders interlaced by rows of single spin-1/2s. In the trimer basis, the model can be made sign-free for any values of $J_{1}$, $J_{2}$,  and $J_{3}$ by extending it to the FFTL model~\cite{Weber2021}, however doing so requires more additional couplings than in the case of the dimer basis.

Going back to the original triangle-square lattice, we may assume that at temperatures $T$ that are large compared to the energy scale of these modifications, the system should behave similarly to the sign-free system. In particular, the average sign should stay finite. At lower temperatures, where these differences become important, the sign will, however, drop towards zero. For the single-spin basis, the energy scale needed to make the model sign-free is that of the intratrimer couplings. For the dimer basis, the scale is the bigger of $|J_1-J_2|$ (to make the model symmetric) and $J$. For the trimer basis it is $J$, although with a larger prefactor since in the extended model, more bonds are added. Thus, unless the mirror symmetry is broken, the dimer basis is superior. After having examined the behavior of the average computational sign, we next consider the thermodynamic behavior of the triangle-square lattice.

\subsection{Thermodynamics}
In the following, we will investigate the thermodynamics of the weakly coupled triangle-square lattice in the regime $T\gtrsim J$, choosing for different points in the phase diagram the best-performing basis according to the preceding analysis. We will commence by looking at the magnetic susceptibility, followed by the specific heat of the system, at the three characteristic sets of intratrimer couplings already considered in the above analysis of the average sign. 

\begin{figure}
	\includegraphics{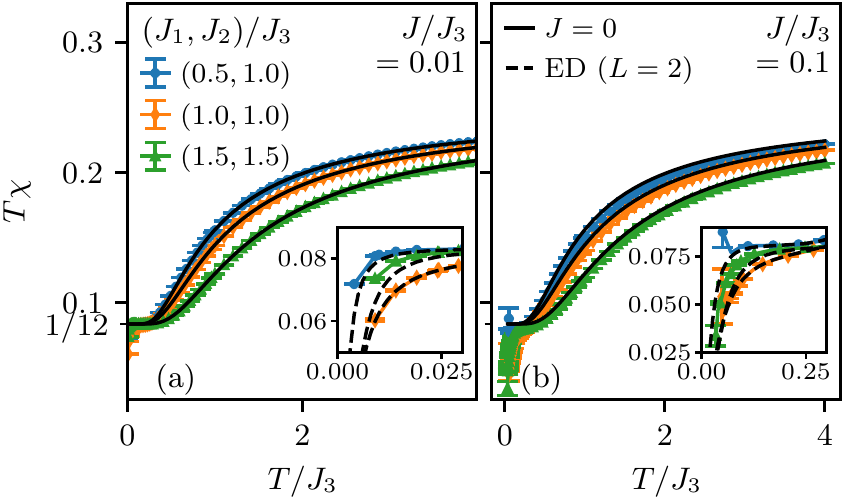}
	\caption{The magnetic susceptibility $\chi$ of the $L=12$ triangle-square lattice, multiplied by the temperature $T$ as a function of temperature for different values of the intra- and intertrimer couplings. The insets show a magnification of the data in the low temperature region. For comparison, the exact solutions for decoupled trimers ($J=0$) and $L=2$ are drawn.}
	\label{fig:triangle_square_chi}
\end{figure}
The magnetic susceptibility $\chi$ (\cref{fig:triangle_square_chi}) shows that at weak intertrimer coupling $J \ll J_1, J_2, J_3$, the physics at high temperatures is well described by an ensemble of decoupled trimers ($J=0$). In this ensemble, there is a crossover from the high-temperature Curie law $T\chi = 1/4$ to a lower-temperature regime, where the spins on each trimer combine into one effective spin-1/2, corresponding to a $T\chi=1/12$ Curie law. At even lower temperatures, the uncoupled-trimer approximation breaks down as the intertrimer interactions couple the effective spin-1/2s, eventually leading to different AFM orders compatible with the square parent lattice of the trimers~\cite{Wessel2001,Wang2001}. This crossover happens at different temperatures, depending on the intratrimer couplings, with $(J_1,J_2)/J_3 = (0.5,1.0)$ being the most robust to interactions out of those studied here. With increasing intertrimer interaction $J$, 
deviations from the decoupled-trimer form are also visible at higher temperatures.
We note that while the exact solution for $L=2$ captures this behavior qualitatively, it significantly differs from the $L=12$ QMC results at low temperatures. This indicates an increased correlation length at low temperatures in the triangle-square lattice. At the same time, this deviation confirms that the QMC approach does perform well beyond the temperature regime of trivial few-trimer physics.

\begin{figure}
	\includegraphics{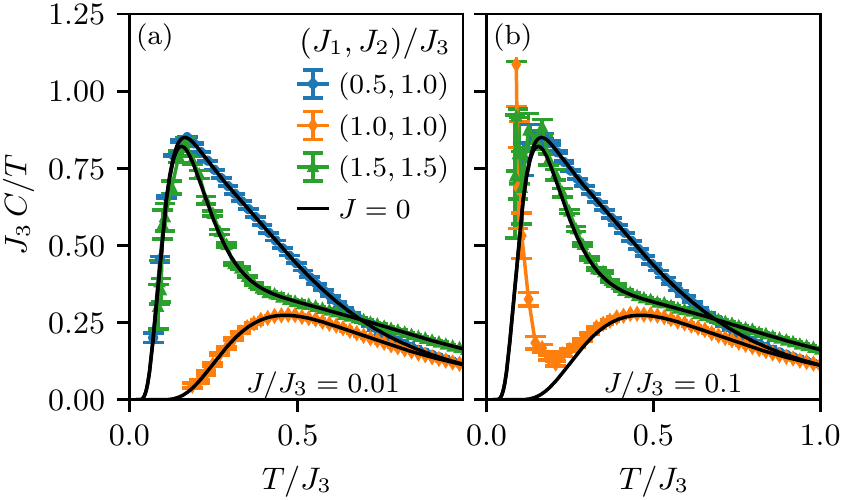}
	\caption{The specific heat $C$ of the $L=12$ triangle-square lattice, divided by the temperature $T$ as a function of temperature for different values of the intra- and intertrimer couplings. Black lines show the exact solution for $J=0$.}
	\label{fig:triangle_square_specheat}
\end{figure}
The specific heat $C$ provides a different perspective on this situation (\cref{fig:triangle_square_specheat}). Here, at high temperatures, the data is again well described by the decoupled case (showing even less deviations than $\chi$ at $J/J_3 = 0.1$). We note that the area under the $C/T$ curves shown in \cref{fig:triangle_square_specheat} corresponds to the released entropy of the model and a deviation from the value $\log 2$ per spin signifies a residual extensive ground state degeneracy. For $J=0$ this is indeed the case as the ground state is a product state of doublets (or quartets in the special case $J_1=J_2=J_3$). At low temperatures, the intertrimer interactions will in general lift this degeneracy leading to an additional release of entropy. For $J/J_3=0.01$, we cannot resolve this regime due to the severity of the sign-problem at such low temperatures. At $J/J_3 = 0.1$, the corresponding temperature scale is higher and we clearly resolve a low-temperature peak in the case $(J_1,J_2)/J_3 = (1,1)$ corresponding to the lifting of the quartet degeneracy.

In summary, in this section, we computed the thermodynamics of the triangle-square lattice, resolving different temperature regimes where different couplings of the Hamiltonian start to play a role. 
In the next step, we will take a closer look at the low-temperature regime and check various predictions of ground-state order in the triangle-square lattice.

\subsection{Ground-state orders}
\label{Sec:TSGSO}
As mentioned in the previous section, at low temperatures, each trimer of the triangle-square lattice forms an effective spin-1/2, or in other words, only the lowest energy doublet contributes to the physics. In a leading-order perturbative expansion in the weak intertrimer coupling $J$, an effective Heisenberg model 
\begin{equation}
	H_\text{eff} =  \sum_{\Delta} J_x^\text{eff} \mathbf{s}_\Delta \cdot \mathbf{s}_{\Delta +\hat{x}} + J_y^\text{eff} \mathbf{s}_\Delta \cdot \mathbf{s}_{\Delta +\hat{y}}
\end{equation}
for these effective spins, denoted $\mathbf{s}_\Delta = P_{\Delta}\,\mathbf{S}_{\Delta}\,P_{\Delta}$, can be derived~\cite{Wang2001}, where $P_{\Delta}$ is a projector to the lowest energy doublet on trimer ${\Delta}$. Because the trimers themselves are arranged on a square parent lattice, this effective low-energy model is no longer frustrated and its ground state phase-diagram is well known to display magnetic order at different wave vectors $\mathbf{Q}$ depending on the signs of $J^\text{eff}_{x/y}$, which in turn depend on the values of $J_{1/2/3}$~\cite{Wessel2001,Wang2001}.

\begin{figure}
	\includegraphics{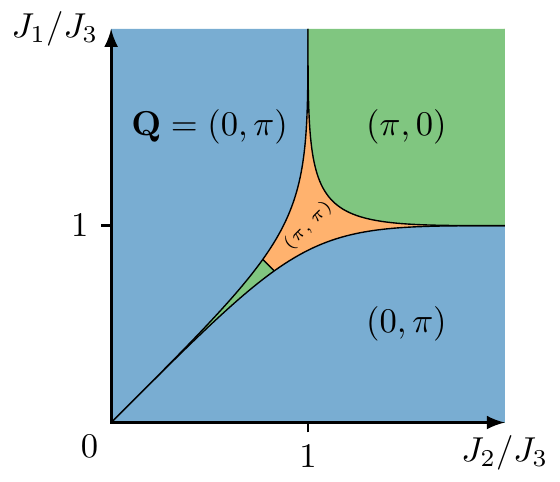}
	\caption{Illustration of the magnetic phase diagram of the triangle-square lattice model obtained from ED in Ref.~\cite{Wessel2001}. Depending on the intratrimer coupling ratios, the effective spin-1/2 degrees of freedom per trimer form magnetic order along different wave vectors $\mathbf{Q}$. The extent of the $(\pi,\pi)$ phase depends on the magnitude of $J$, shrinking to the singular point $J_1=J_2=J_3$ as $J\rightarrow 0$; the illustration corresponds to $J\approx 0.1 J_3$. }
	\label{fig:triangle_square_phasediag}
\end{figure}

The case $J_1=J_2=J_3$ forms an exception in this analysis since the low-energy subspace consists of two degenerate SU(2) doublets. In addition to an effective  SU(2) spin $\mathbf{s}_\Delta$, this gives rise to a pseudospin low-energy degree of freedom $\boldsymbol{\tau}$ distinguishing the two doublets. The effective Hamiltonian then becomes
\begin{equation}
	\label{eq:ts_degenerate_ham}
	H_\text{eff}' = J \sum_{\Delta} \mathbf{s}_\Delta \cdot \mathbf{s}_{\Delta +\hat{x}} A_{\Delta, \Delta+\hat{x}} + \mathbf{s}_\Delta \cdot \mathbf{s}_{\Delta +\hat{y}} A_{\Delta, \Delta+\hat{y}},
\end{equation}
where the operators $A_{\Delta,\Delta'}$ act on the pseudospins (given in detail in Ref.~\cite{Wang2001} and further below). The ground state of this model is not well known, but ED  and mean-field theory suggest the formation of $(\pi,\pi)$ order around $J_1=J_2=J_3$. A sketch of the complete phase diagram as obtained in Ref.~\cite{Wessel2001} is shown in \cref{fig:triangle_square_phasediag}.

In the following, we will check the validity of this phase diagram for finite but small intertrimer couplings. To this end, we compute the magnetic structure factor,
\begin{equation}
	S(\mathbf{Q}) = \frac{1}{L^2} \sum_{\Delta, \Delta'} e^{i \mathbf{Q}\cdot(\mathbf{R}_\Delta-\mathbf{R}_{\Delta'})} \braket{\mathbf{S}_{\Delta} \cdot \mathbf{S}_{\Delta'}},
\end{equation}
with the unit cell positions $\mathbf{R}_\Delta$, while simultaneously scaling the system size $L$ and the temperature $T=J_3/2L$ to resolve the onset of the different magnetic ground state orders. Since the sign problem is least severe along the symmetric line $J_1=J_2$, we concentrate on this case in the following analysis.

\begin{figure}
	\includegraphics{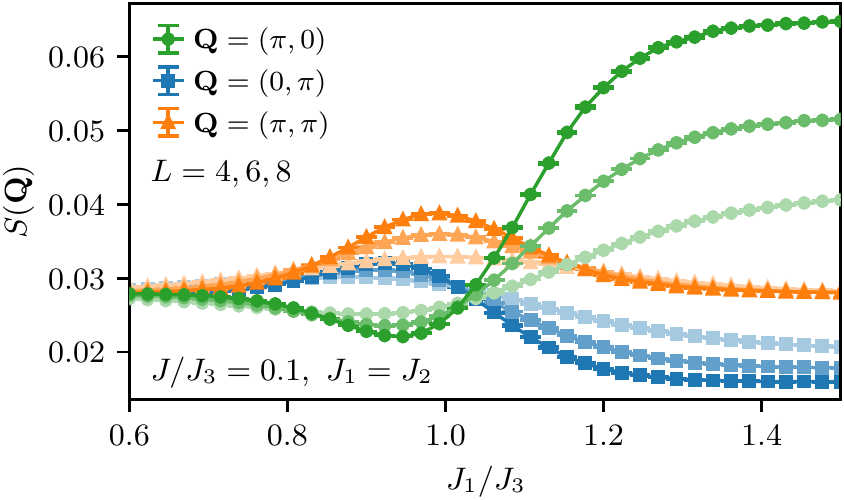}	\caption{Finite-size and temperature scaling ($T=J_3/2L$) of the magnetic structure factor $S(\mathbf{Q})$ of the triangle-square lattice along the symmetric line ($J_1=J_2$, $J/J_3=0.1$, see \cref{fig:triangle_square_phasediag}). The system size increases with the opacity of the lines. }
	\label{fig:triangle_square_strucfac}
\end{figure}

A scan of the structure factor at different possible ordering wave vectors (\cref{fig:triangle_square_strucfac}) reveals that along the symmetric line, the $(\pi,0)$ and $(\pi,\pi)$ orders are most prevalent, with the former dominating at high $J_1=J_2$ and the latter taking over in the regime around $J_1=J_2=J_3$. At lower $J_1=J_2$, the competition between different orders is very close. However, out of the three structure factors, the one belonging to $(\pi,0)$ is the only one growing with system size in this region, suggesting that it will dominate in the thermodynamic limit.
These findings are in agreement with the phase diagram obtained using ED, perturbation theory and mean-field theory. In particular, the close competition at low $J_1=J_2$ can be explained by a weak effective coupling $J_y^\text{eff}$ that only appears at higher orders perturbatively in the intertrimer coupling $J$ \cite{Wang2001}.

As we saw, most of the phase diagram of the triangle-square lattice can be understood in terms of the effective spin-1/2 degrees of freedom formed by each trimer. So far, however, we have not considered the pseudospin degree of freedom at the point $J_1=J_2=J_3$. This degree of freedom has a similar role as the third quantum number we had to add in the trimer basis in \cref{sec:basis} in that it lives in the subspace of fixed $l_\Delta$ and $m_\Delta$. In  \cref{sec:basis}, we used the eigenbasis of the operator $(\mathbf{S}_{\Delta,1}+\mathbf{S}_{\Delta,2})^2$ to map out this part of the trimer Hilbert space while retaining a real basis. Without this constraint, it is however often useful to use a basis that makes the symmetry of the $J_1=J_2=J_3$ trimer explicit. One such basis is the eigenbasis of the chirality operator~\cite{Subrahmanyam95}
\begin{equation}
	\tau^z_\Delta := \frac{\sqrt{3}}{4} \mathbf{S}_{\Delta,1} \cdot \left(\mathbf{S}_{\Delta,2} \times \mathbf{S}_{\Delta,3}\right) =: \ketbra{R}{R} - \ketbra{L}{L},
\end{equation}
with
\begin{align}
	\ket{L,\ua} &= \frac{1}{\sqrt{3}} \left(\ket{\ua\ua\da} + \omega \ket{\ua\da\ua} + \omega^* \ket{\da\ua\ua}\right),\\
	\ket{L,\da} &= \frac{1}{\sqrt{3}} \left(\ket{\da\da\ua} + \omega \ket{\da\ua\da} + \omega^* \ket{\ua\da\da}\right),\\
	\ket{R,\ua} &= \frac{1}{\sqrt{3}} \left(\ket{\ua\ua\da} + \omega^* \ket{\ua\da\ua} + \omega \ket{\da\ua\ua}\right),\\
	\ket{R,\da} &= \frac{1}{\sqrt{3}} \left(\ket{\da\da\ua} + \omega^* \ket{\da\ua\da} + \omega \ket{\ua\da\da}\right),
\end{align}
and $\omega = e^{i 2\pi/3}$. Writing the effective Hamiltonian of \cref{eq:ts_degenerate_ham} in this basis yields \cite{Wang2001}
\begin{align}
	H_\text{eff} = \frac{J}{9} \sum_{\Delta} & A^l_{\Delta}A^r_{\Delta+\hat{x}}\,\mathbf{S}_\Delta \cdot \mathbf{S}_{\Delta+\hat{x}}\nonumber\\
	+ &A^d_{\Delta}A^u_{\Delta+\hat{y}}\, \mathbf{S}_\Delta \cdot \mathbf{S}_{\Delta+\hat{y}},
	\label{eq:ff:ts_heff}
\end{align}
where \footnote{Because our spin numbering convention differs from Ref.~\cite{Wang2001}, we pick up an extra phase $\omega$ here, simplifying the Hamiltonian slightly.}
\begin{align}
	A^l_{\Delta} &= 1-2\omega^*\tau^+_\Delta-2\omega \tau^-_\Delta,\nonumber\\
	&= 1 + \tau_\Delta^x + \sqrt{3} \tau_\Delta^y,& \\
	A^r_\Delta &= 1-2 \omega \tau^+_{\Delta} - 2 \omega^* \tau^{-}_{\Delta},\nonumber\\
	&=1+\tau_\Delta^x - \sqrt{3} \tau_\Delta^y\\
	A^d_{\Delta} &= 1-2 \tau^x_\Delta,\\A^u_{\Delta} &= 2+2 \tau^x_\Delta.
\end{align}
act on the pseudospin degree of freedom via the operators $\tau^+ = \ketbra{L}{R}$ and $\tau^- = (\tau^+)^\dagger$. Note that the shape of the $A$ operators is strongly constrained by the symmetries of the triangle-square lattice. The horizontal mirror symmetry $\Sigma$, 
represented by 
$\tau^x$ (skipping the trimer index $\Delta$) on each trimer implies 
\begin{equation}
	\Sigma A^l \Sigma^{-1} = A^r, \quad \Sigma A^d \Sigma^{-1} = A^d, \quad \Sigma A^u \Sigma^{-1} = A^u.
\end{equation}
Similarly, the time-reversal symmetry $\Theta$, given by $K \tau^x$, where $K$ is the complex conjugation operator in the chirality basis, leads to
\begin{align}
	\Theta A^l \Theta^{-1} &= A^l,& \Theta A^r \Theta^{-1} &= A^r,\\
	\Theta A^d \Theta^{-1} &= A^d, & \Theta A^u \Theta^{-1} &= A^u.
\end{align}

At low temperatures, the pseudospin degree of freedom may spontaneously break either of these symmetries, forming a chiral or lattice nematic state. The former was indeed proposed based on mean-field theory results for $H_\text{eff}'$ \cite{Wang2001}. In contrast to the magnetic order which constitutes the breaking of a continuous symmetry, the pseudospin can break its discrete symmetries also at finite temperature.

To check these possibilities, we compute the correlation functions of the chirality pseudospin $\tau^z$, which is odd under time reversal, 
\begin{equation}
	C^z(\mathbf{r}) = \braket{\tau^z_0 \tau^z_{\mathbf{r}}},
\end{equation}
and the operator $\tau^y = \frac{2}{\sqrt{3}} (\mathbf{S}_1 - \mathbf{S}_2)\cdot \mathbf{S}_3$, which is odd under horizontal reflections,
\begin{equation}
	C^y(\mathbf{r}) = \braket{\tau^y_0 \tau^y_{\mathbf{r}}}.
\end{equation}
Since we cannot directly simulate in the chirality basis, both of these correlation functions contain operators that are off-diagonal in our computational basis. Nevertheless, it is possible to compute them on-the-fly during the loop update. In \cref{sec:loopmeas} we show how such on-the-fly measurements can be performed within the  SSE loop update also in the presence of a sign problem.

Having identified $\tau^z$ and $\tau^y$ in terms of constituent spin operators, we note that $\tau^x = (\mathbf{S}_1 + \mathbf{S}_2)^2 - 1$. This highlights that the chirality basis and the computational “$l_{\Delta,12}$” trimer basis introduced in \cref{sec:basis} are related by a simple unitary pseudospin rotation. Consequently, in the $l_{\Delta,12}$-trimer basis, the correlation functions $C^z$ and $C^y$ can still be expressed by simple (albeit different) Pauli matrices,
\begin{align}
	\tau^z &= \ketbra{1/2}{1/2}_{l_\Delta}\otimes \mathbf{1}_{m_\Delta} \otimes i(\ketbra{1}{0} - \ketbra{0}{1})_{l_{\Delta,12}},\\
	\tau^y &= \ketbra{1/2}{1/2}_{l_\Delta}\otimes \mathbf{1}_{m_\Delta} \otimes (\ketbra{1}{0} + \ketbra{0}{1})_{l_{\Delta,12}}.
\end{align}
For this reason, we prefer the trimer basis over the dimer computational basis for this calculation even though it has a slightly reduced average sign. In the dimer basis, calculating $C^z$ and $C^y$ in a similar way is in principle possible, but with much more complicated matrix elements due to the split of the Hilbert space of $\mathbf{S}_{\Delta,1}$, $\mathbf{S}_{\Delta,2}$, and $\mathbf{S}_{\Delta,3}$ constituting the spin chirality into two distinct computational cluster-basis sites.

\begin{figure}
	\includegraphics{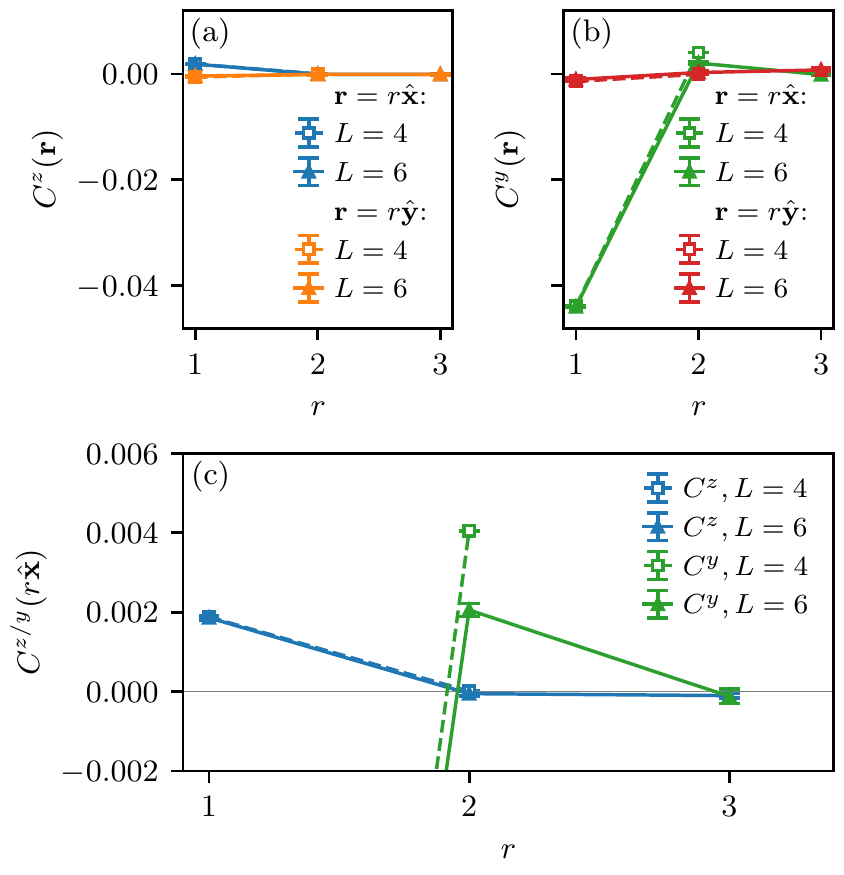}
	\caption{Chirality correlation functions $C^z$ and $C^y$ for the triangle-square lattice for $J_1=J_2=J_3$, $J/J_3=0.1$ at $T/J_3=0.05$. Panels (a) and (b) show the respective correlations in both the $x$ and $y$ lattice directions while panel (c) compares them along the dominant $x$ direction.}
	\label{fig:triangle_chiral}
\end{figure}

As seen from the QMC data shown for  $C^z(\mathbf{r})$ (top left panel a) and $C^y(\mathbf{r})$ (top right panel b)  in \cref{fig:triangle_chiral}, we observe a rapid decay of these correlations along both lattice directions, i.e., for $\mathbf{r} \propto \hat{\mathbf{x}}$ ($x$ direction)  and $\mathbf{r}\propto\hat{\mathbf{y}}$ ($y$ direction). The chirality  correlations exhibit a finite value only among nearest neighbor trimers, and are essentially zero beyond $r>2$ (the value of $C^y(\mathbf{r})$ at $\mathbf{r}=2\,\hat{\mathbf{x}}$ is also  suppressed with increasing system size $L$, cf. the bottom panel c). We therefore do not obtain indication for the presence of any further symmetry breaking induced by the chirality degree of freedom~\cite{Wang2001} within the accessible temperature regime of our QMC approach.

\section{Breathing kagome lattice}
After having considered a system of coupled trimers, for which the superlattice of trimers is bipartite, we next consider the case of the (breathing) kagome lattice, which is formed by a (non-bipartite) triangular lattice of coupled trimers. In this system there is thus an additional source of magnetic frustration beyond the one introduced by the antiferromagnetic  intratrimer couplings. 
We again consider finite-size systems with periodic boundary  conditions, and also in this case the number of spins relates via $N=3L^3$ to the linear system size $L$  of the triangular lattice. 
\label{sec:bk}
\subsection{Sign}
\begin{figure}
	\includegraphics{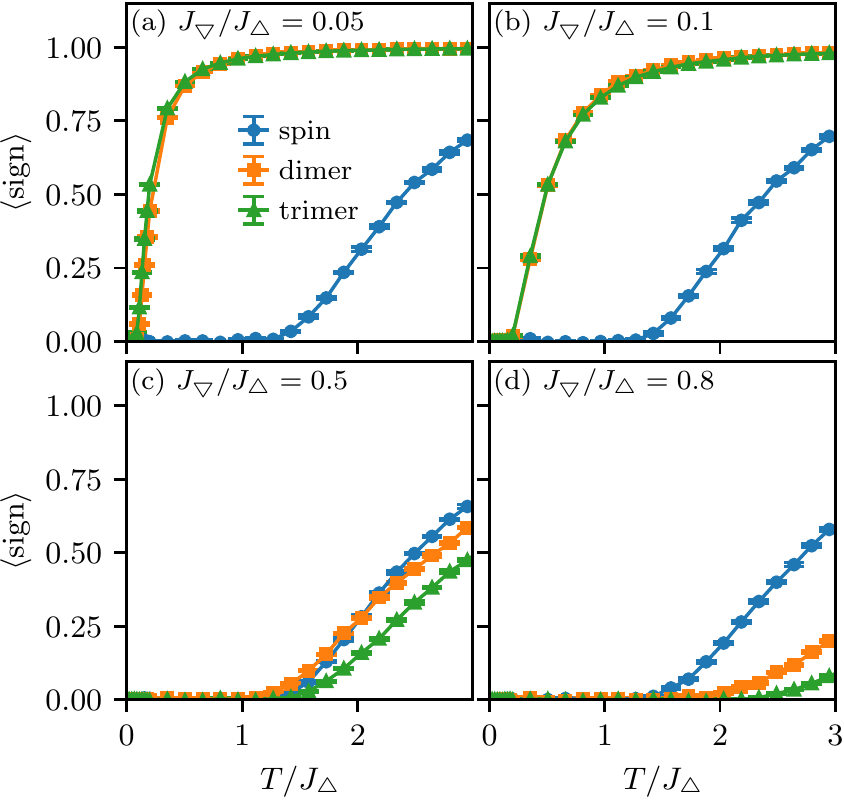}
	\caption{Average configuration sign, $\langle \mathrm{sign}\rangle$, of the $L=12$ breathing kagome lattice in different computational bases for different breathing distortions $J_\bigtriangledown/J_\triangle$.}
	\label{fig:kagome_sign}
\end{figure}
For the kagome lattice with breathing distortion, we also start by comparing the average configuration sign for the computational bases introduced in \cref{sec:basis} (\cref{fig:kagome_sign}). For weak intertrimer couplings $J_\bigtriangledown$, the dimer and trimer bases again strongly outperform the single-spin basis retaining a finite sign down to $T\sim J_\bigtriangledown$, while at stronger couplings this advantage fades away. In contrast to the triangle-square lattice, however, at weak intertrimer couplings, the sign is nearly identical between the dimer and trimer basis for $J_\bigtriangledown/J_\triangle \lesssim 0.1$, with a slight advantage of the trimer basis in the low-$T$ regime. At larger $J_\bigtriangledown$, the dimer basis has a slightly higher average sign, but eventually falls below the sign of the single-spin basis.
Therefore, moving forward, we will focus on the weak coupling regime, i.e. the regime of high breathing distortion, and use the trimer computational basis.
\subsection{Thermodynamics}
\begin{figure}
	\includegraphics{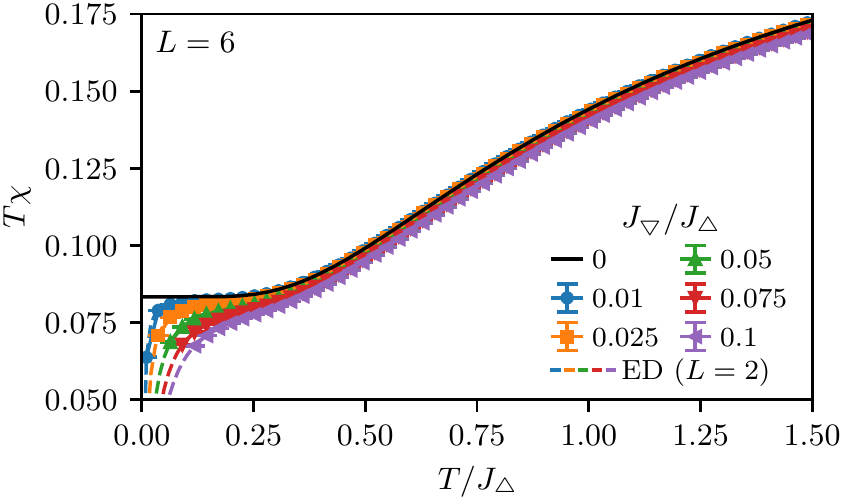}
	\caption{The magnetic susceptibility $\chi$ of the breathing kagome lattice multiplied by the temperature $T$ as a function of $T$ for different breathing distortions $J_\bigtriangledown/J_\triangle$. For $J_\bigtriangledown=0$, the ED solution for decoupled trimers is shown. For finite $J_\bigtriangledown$, the dashed lines show $L=2$ ED data.}
	\label{fig:kagome_chi}
\end{figure}

\begin{figure}
	\includegraphics{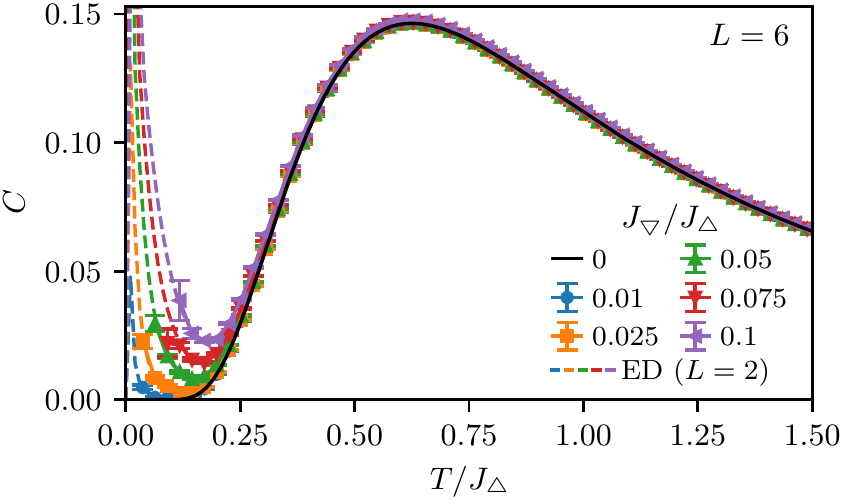}
	\caption{The specific heat $C$ of the breathing kagome lattice as a function of temperature $T$ for different breathing distortions $J_\bigtriangledown/J_\triangle$. For $J_\bigtriangledown = 0$ the ED solution for decoupled trimers is shown. For finite $J_\bigtriangledown$, the dashed lines show $L=2$ ED data.}
	\label{fig:kagome_specheat}
\end{figure}
In analogy to the triangle-square lattice, we compute the magnetic susceptibility $\chi$ (\cref{fig:kagome_chi}) and the specific heat $C$ (\cref{fig:kagome_specheat}). At high temperatures and weak intertrimer couplings, these observables converge to the same decoupled trimer limit as the triangle-square lattice. In the susceptibility (\cref{fig:kagome_chi}), the effective spin-1/2 plateau at low temperatures again vanishes with increasing intertrimer interactions. The specific heat shows the onset of a low temperature peak associated with the lifting of degeneracies by the intertrimer interactions. In general, due to the sign problem it is challenging to reach the low-temperature regime $T\sim J_\bigtriangledown$, but in our accessible temperature range both the susceptibility and the specific heat are well described by ED data for $L=2$. This is in contrast to the triangle-square lattice for which at similar temperatures, significant deviations from the $L=2$ data were found due to the onset of magnetic correlations.

\begin{figure}
	\includegraphics{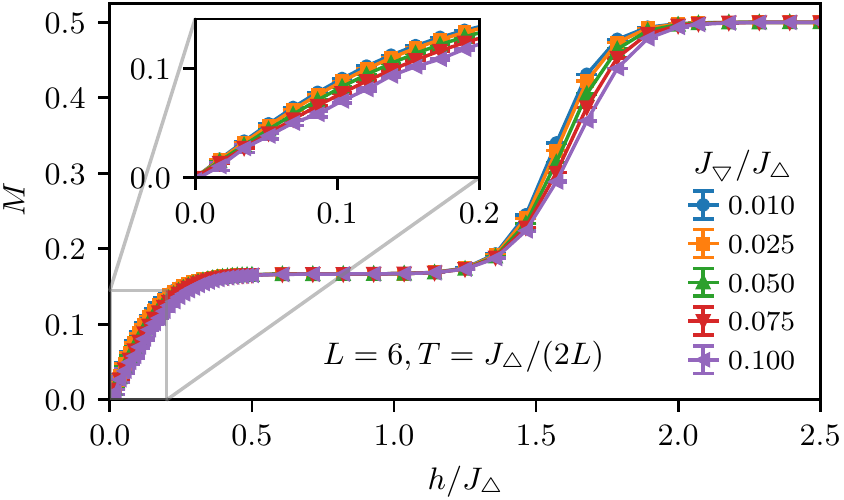}
	\caption{Magnetization per site $M$ of the breathing kagome lattice as a function of magnetic field $h$ for different breathing distortions $J_\bigtriangledown/J_\triangle$, taken at $T=0.083J_\triangle$.}
	\label{fig:kagome_mag}
\end{figure}
In addition to these observables, we furthermore computed the magnetization (per site) $M=\frac{1}{M}\sum_i S^z_i$ at a finite magnetic field $h$ introduced to the Hamiltonian,
\begin{equation}
H' = H - h \sum_{\Delta} S_\Delta^z\, ,
\end{equation}
which shows two extended plateaus (\cref{fig:kagome_mag}) corresponding to the magnetizations per trimer $\braket{m_\Delta}=1/2$ and $\braket{m_\Delta}=3/2$, respectively~\cite{Honecker11}. At the boundaries of these plateaus, the  competition between different magnetization sectors renders the energetic contributions due to the small intertrimer coupling more important. In these regions the magnetization displays smooth crossovers, where at our lowest accessible temperature of around $T/J_\triangle = 0.083$, we do not resolve signs of further plateaus. In general, increasing $J_\bigtriangledown$ at fixed $T/J_\triangle$ leads to stronger frustration and thus to an overall decrease of the magnetization.

Previous works have found a quantum phase transition from a nematic state breaking the lattice rotation symmetry to a spin liquid state at weak intertrimer coupling in the range of $J_\bigtriangledown/J_\triangle \approx 0.05$ \cite{Repellin2017}. While we see no signatures of such a transition in the thermodynamic behavior presented until now, measuring the associated nematicity order parameter and its correlation function is in principle possible within our approach, allowing for the direct detection of such a phase, if it persists to high enough temperatures. In \cref{sec:nematicity}, we show the absence of a lattice nematic state for $T/J_\triangle \gtrsim 0.083$, limiting a finite-temperature lattice nematic phase, if it persits, to lower temperatures than those accessible to our QMC approach. 

\section{Conclusion}
\label{sec:conc}
In this article we computed the thermodynamics of two models of weakly coupled Heisenberg trimers using quantum Monte Carlo simulations in different cluster bases. In both the triangle-square lattice and the breathing kagome lattice, these cluster bases allowed us to resolve temperatures down to the order of the weak intertrimer coupling, greatly outperforming the single-spin basis.

For the triangle-square lattice, we further picked up the signatures of different magnetic orders expected to form in the ground state. We also showed that correlations related to the additional chirality degree of freedom for the case of equal intratrimer couplings can be efficiently accessed using the trimer basis. However, no additional  symmetry breaking as observed within the accessibly  temperature range. 

The thermodynamic data for the  breathing kagome lattice was found to be well-described by a simple $L=2$ cluster throughout our accessible temperature range. In particular, we observe the presence of a two-peak structure in the specific heat in both the lattice-nematic and the quantum spin liquid regime of the breathing kagome lattice. Whereas some studies also reported a two-peak structure in the undistored kagome lattice ($J_\bigtriangledown=J_\triangle$)~\cite{misguich05,Shimokawa16}, recent work concludes instead in favor of a single peak with a pronounced low-$T$ shoulder in the undistored kagome lattice~\cite{Sugiura13,Chen18,Schnack18}. It would therefore be interesting to further examine the evolution of the low-temperature specific heat upon increasing $J_\bigtriangledown$ from the strong breathing regime that we can access by QMC into the weak breathing region, including the undistored limit. 

\begin{acknowledgments}
We thank Andreas Honecker and J\"urgen Schnack for useful discussions. 
We acknowledge the support of the Deutsche Forschungsgemeinschaft (DFG, German Research Foundation) through Grant No. WE/3649/4- 2 of the program FOR 1807 and through project RTG 1995, and the IT Center at RWTH Aachen University and the JSC J\"ulich for access to computing time through JARA-HPC.
\end{acknowledgments}

\appendix
\section{Measuring correlation functions}
\label{sec:loopmeas}
In the $S^z$ basis formulation of the SSE it is possible to measure the offdiagonal spin correlation functions $\braket{S^+_i S^-_j}$ by treating the (directed) loop update as an extended ensemble where theses operators are part of the configuration~\cite{Dorneich2001,Alet2005}. 
This approach can be directly generalized to  correlations $\braket{A_I B_J}$ where $A_I$ and $B_I$ are local operators on cell (i.e., trimer) $I$: In particular, these measurements can also be performed in the presence of signed configurations, as we will show in the following, based on the notation introduced in Ref.~\cite{Weber2021}.
If we choose a loop action (out of a total of $N_{\text{actions}}$) and the  entry cell (out of a total of $N_{\text{cells}}$) at random, the extended ensemble takes the form
\begin{align}
	&Z_{\text{ext}} = \sum_{\substack{n,\{b_n\},\sigma_0\\ a_h, p_h, a_t, p_t}}  \frac{1}{N_{\text{actions}} N_{\text{cells}}} \Big|\frac{(-\beta)^n}{n!} \\ \!\times&\braket{\sigma_0|h_{b_1}\cdots|\sigma_{p_h}} \braket{a_h(\sigma_{p_h})| \cdots|\sigma_{p_t}}\braket{a_t(\sigma_{p_t})| \cdots h_{b_n}|\sigma_0}\Big|.\nonumber
\end{align}
Here, in addition to the expansion order $n$, operator string $\{b_n\}$ and state $\sigma_0$  the configuration contains the position $p$ and action $a$ of the loop's head and tail. Furthermore, the system Hamiltonian $H$ has been decomposed in terms of  operators $h_b$ that each connect two cells~\cite{Weber2021} (they correspond to the bond operators in the conventional site-basis formulation of the SSE~\cite{Sandvik1991}). The $a_t$, $p_t$, $a_h$, and $p_h$ determine the position and type of two discontinuities that appear in the string of nonbranching operators~\cite{Dorneich2001,Alet2005}.

We want to measure equal-time correlation functions between general state, i.e., 
\begin{equation}
	C_{x_i y_i, x_j y_j} =\frac{\Tr\left[\ket{x_i}\!\bra{y_i}\otimes\ket{x_j}\!\bra{y_j} e^{-\beta H}\right]}{Z}.
\end{equation}
If the local states $x \neq y$, there is a loop action $a$ so that $y = a(x)$ and we can express the correlations as an expectation value in the extended ensemble
\begin{equation}
	C_{x_i y_i, x_j y_j} = N_\text{actions} N_\text{cells} \frac{\braket{\text{sign} \times P(a_h, p_h, a_t, p_t)}_\text{ext}}{\braket{\text{sign} \times Q(a_h, p_h, a_t, p_t)}_\text{ext}}
\end{equation}
with two projectors
\begin{align}
	P(a_h,p_h, a_t,p_t) &=\begin{cases}1, &\text{singularities match } x_i y_i, x_j y_j\\ 0, & \text{else}\end{cases}\\
	Q(a_h,p_h, a_t,p_t) &=\begin{cases}1, &a_h = a_t^{-1} \text{ and } p_h = p_t\\ 0, & \text{else}\end{cases}
\end{align}
i.e.,  $Q$ filters out closed-loop configurations that are also part of the regular ensemble.
Therefore, the denominator can be simplified to
\begin{equation}
	\braket{\text{sign} \cdot Q(a_h, p_h, a_t, p_t)}_\text{ext} = \braket{\text{sign}} \braket{Q(a_h, p_h, a_t, p_t)}_\text{ext}.
\end{equation}
In the numerator, we write the sign as a difference of projectors $P_+$ ($P_-$) on the positive (negative) signed configurations
\begin{align}
	\braket{\text{sign} \cdot P(a_h, p_h, a_t, p_t)}_\text{ext} &= \braket{P_+ P(a_h, p_h, a_t, p_t)}_\text{ext} \\ &-\braket{P_- P(a_h, p_h, a_t, p_t)}_\text{ext}.\nonumber
\end{align}
Analogously to the usual case~\cite{Alet2005,Dorneich2001}, we can write the ratio of the two probabilities $\braket{P}$ and $\braket{Q}$ as mean counts per loop. Thus, 
\begin{align}
	C_{x_i y_i, x_j y_j} &= N_\text{actions} N_\text{cells}\\
	&\times \frac{\braket{n_+(a_h, p_h, a_t, p_t) - n_-(a_h, p_h, a_t, p_t)}}{\braket{\text{sign}}}\nonumber\\
	&= \frac{\braket{n(a_h, p_h, a_t, p_t)}}{\braket{\text{sign}}}
\end{align}
where $n(a_h, p_h, a_t, p_t)$ is the \textit{signed} count of loop steps with $P(a_h, p_h, a_t, p_t)=1$.
Arbitrary correlation functions of cell-local operators can now be written as sum of the $C_{x_i y_i, x_j y_j}$. If they are offdiagonal ($(x_i, y_i) \neq (x_j, y_j)$), they can be measured as described in this section. If they are diagonal, they can be measured like every other diagonal observable~\cite{Sandvik1991,Sandvik1992}.

\section{Lattice nematic response in the breathing kagome lattice}
\label{sec:nematicity}

\begin{figure}[t!]
	\includegraphics{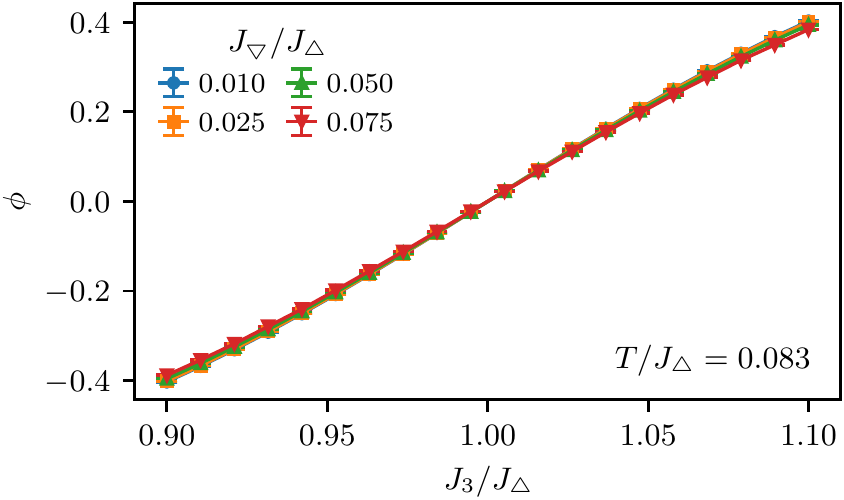}
	\caption{The nematic order parameter $\phi$ in the $L=6$ breathing kagome lattice as a function of the intertrimer coupling $J_3$ detuned from the C\textsubscript{3}-symmetric point $J_3=J_\triangle$ for different intertrimer couplings $J_\bigtriangledown$. }
	\label{fig:nematicity}
\end{figure}

At low $J_\bigtriangledown/J_\triangle\lesssim 0.05$ and $T=0$, the kagome lattice AFM was previously found to spontaneously break the C\textsubscript{3} lattice rotational symmetry while leaving the translational, spin rotational, and a lattice reflection symmetry intact~\cite{Repellin2017,Jahromi2020}. Such a “lattice nematic” phase is expected to show a singular response to a small C\textsubscript{3} breaking perturbation, which can be detected, e.g., using an order parameter defined as
\begin{equation}
	\phi = \frac{1}{L^2}\sum_\Delta \sum_{n=1}^3 e^{2 \pi i (n-1)/3} \braket{\mathbf{S}_{\Delta,n} \cdot \mathbf{S}_{\Delta,n+1}}.
\end{equation}
In the C\textsubscript{3}-symmetric case, this quantity is exactly zero.

Previous studies have focused on the ground state phase-diagram, but since the lattice nematic phase breaks a discrete symmetry it may persist up to a finite temperature $T_c$. As $J_\bigtriangledown$ approaches zero, at any fixed temperature, the partition function of the model should approach the one of simple decoupled trimers, in agreement with our findings in the main text. Decoupled trimers cannot, of course, form lattice nematic order, so the critical temperature $T_c$ of the nematic transition is expected to vanish with $J_\bigtriangledown$. Nevertheless, at finite $J_\bigtriangledown$, a finite $T_c$ may be observable in the temperature range accessible with our method, which is what we investigate in this section.

We will introduce a small symmetry-breaking perturbation by detuning one of the intratrimer couplings, denoted as $J_3$ (in analogy to the triangle-square lattice notation) while leaving the other two intratrimer couplings at the value $J_\triangle$. We can then monitor the evolution of the order parameter upon tuning across the isotropic point $J_3=J_\triangle$. Using the residual lattice reflection symmetry in this setup, the order parameter $\phi$ can be expressed as
\begin{equation}
	\phi = \frac{1}{L^2} \left(\Braket{\sum_{\Delta} l_{\Delta,12}} - \Braket{\sum_{\Delta} l_{\Delta,23}}\right),
\end{equation}
Here, we computed the two terms as diagonal observables in two separate simulations in appropriately rotated trimer bases.

The results for $\phi$ even at our lowest accessible temperature show a smooth dependence both on $J_3/J_\triangle$ and $J_\bigtriangledown/J_\triangle=0.010 - 0.075$ (\cref{fig:nematicity}). In particular, for all values of $J_\bigtriangledown/J_\triangle$, the data in Fig.~\cref{fig:nematicity} smoothly tend towards zero upon approaching the isotropic point 
$J_3=J_\triangle$. We therefore conclude that a finite-temperature lattice nematic phase, if it persists, is limited to lower temperatures than accessible to our QMC approach. 

\bibliography{paper.bib}

\begin{thebibliography}{44}%
\makeatletter
\providecommand \@ifxundefined [1]{%
 \@ifx{#1\undefined}
}%
\providecommand \@ifnum [1]{%
 \ifnum #1\expandafter \@firstoftwo
 \else \expandafter \@secondoftwo
 \fi
}%
\providecommand \@ifx [1]{%
 \ifx #1\expandafter \@firstoftwo
 \else \expandafter \@secondoftwo
 \fi
}%
\providecommand \natexlab [1]{#1}%
\providecommand \enquote  [1]{``#1''}%
\providecommand \bibnamefont  [1]{#1}%
\providecommand \bibfnamefont [1]{#1}%
\providecommand \citenamefont [1]{#1}%
\providecommand \href@noop [0]{\@secondoftwo}%
\providecommand \href [0]{\begingroup \@sanitize@url \@href}%
\providecommand \@href[1]{\@@startlink{#1}\@@href}%
\providecommand \@@href[1]{\endgroup#1\@@endlink}%
\providecommand \@sanitize@url [0]{\catcode `\\12\catcode `\$12\catcode
  `\&12\catcode `\#12\catcode `\^12\catcode `\_12\catcode `\%12\relax}%
\providecommand \@@startlink[1]{}%
\providecommand \@@endlink[0]{}%
\providecommand \url  [0]{\begingroup\@sanitize@url \@url }%
\providecommand \@url [1]{\endgroup\@href {#1}{\urlprefix }}%
\providecommand \urlprefix  [0]{URL }%
\providecommand \Eprint [0]{\href }%
\providecommand \doibase [0]{http://dx.doi.org/}%
\providecommand \selectlanguage [0]{\@gobble}%
\providecommand \bibinfo  [0]{\@secondoftwo}%
\providecommand \bibfield  [0]{\@secondoftwo}%
\providecommand \translation [1]{[#1]}%
\providecommand \BibitemOpen [0]{}%
\providecommand \bibitemStop [0]{}%
\providecommand \bibitemNoStop [0]{.\EOS\space}%
\providecommand \EOS [0]{\spacefactor3000\relax}%
\providecommand \BibitemShut  [1]{\csname bibitem#1\endcsname}%
\let\auto@bib@innerbib\@empty
\bibitem [{\citenamefont {Starykh}(2015)}]{Starykh2015}%
  \BibitemOpen
  \bibfield  {author} {\bibinfo {author} {\bibfnamefont {O.~A.}\ \bibnamefont
  {Starykh}},\ }\href {\doibase 10.1088/0034-4885/78/5/052502} {\bibfield
  {journal} {\bibinfo  {journal} {Rep. Prog. Phys.}\ }\textbf {\bibinfo
  {volume} {78}},\ \bibinfo {pages} {052502} (\bibinfo {year}
  {2015})}\BibitemShut {NoStop}%
\bibitem [{\citenamefont {Savary}\ and\ \citenamefont
  {Balents}(2016)}]{Savary2016}%
  \BibitemOpen
  \bibfield  {author} {\bibinfo {author} {\bibfnamefont {L.}~\bibnamefont
  {Savary}}\ and\ \bibinfo {author} {\bibfnamefont {L.}~\bibnamefont
  {Balents}},\ }\href {\doibase 10.1088/0034-4885/80/1/016502} {\bibfield
  {journal} {\bibinfo  {journal} {Rep. Prog. Phys.}\ }\textbf {\bibinfo
  {volume} {80}},\ \bibinfo {pages} {016502} (\bibinfo {year}
  {2016})}\BibitemShut {NoStop}%
\bibitem [{\citenamefont {Yan}\ \emph {et~al.}(2011{\natexlab{a}})\citenamefont
  {Yan}, \citenamefont {Huse},\ and\ \citenamefont {White}}]{Yan2011}%
  \BibitemOpen
  \bibfield  {author} {\bibinfo {author} {\bibfnamefont {S.}~\bibnamefont
  {Yan}}, \bibinfo {author} {\bibfnamefont {D.~A.}\ \bibnamefont {Huse}}, \
  and\ \bibinfo {author} {\bibfnamefont {S.~R.}\ \bibnamefont {White}},\ }\href
  {\doibase 10.1126/science.1201080} {\bibfield  {journal} {\bibinfo  {journal}
  {Science}\ }\textbf {\bibinfo {volume} {332}},\ \bibinfo {pages} {1173}
  (\bibinfo {year} {2011}{\natexlab{a}})}\BibitemShut {NoStop}%
\bibitem [{\citenamefont {Depenbrock}\ \emph {et~al.}(2012)\citenamefont
  {Depenbrock}, \citenamefont {McCulloch},\ and\ \citenamefont
  {Schollw\"ock}}]{Depenbrock2012}%
  \BibitemOpen
  \bibfield  {author} {\bibinfo {author} {\bibfnamefont {S.}~\bibnamefont
  {Depenbrock}}, \bibinfo {author} {\bibfnamefont {I.~P.}\ \bibnamefont
  {McCulloch}}, \ and\ \bibinfo {author} {\bibfnamefont {U.}~\bibnamefont
  {Schollw\"ock}},\ }\href {\doibase 10.1103/PhysRevLett.109.067201} {\bibfield
   {journal} {\bibinfo  {journal} {Phys. Rev. Lett.}\ }\textbf {\bibinfo
  {volume} {109}},\ \bibinfo {pages} {067201} (\bibinfo {year}
  {2012})}\BibitemShut {NoStop}%
\bibitem [{\citenamefont {Jiang}\ \emph {et~al.}(2012)\citenamefont {Jiang},
  \citenamefont {Wang},\ and\ \citenamefont {Balents}}]{Jiang2012}%
  \BibitemOpen
  \bibfield  {author} {\bibinfo {author} {\bibfnamefont {H.-C.}\ \bibnamefont
  {Jiang}}, \bibinfo {author} {\bibfnamefont {Z.}~\bibnamefont {Wang}}, \ and\
  \bibinfo {author} {\bibfnamefont {L.}~\bibnamefont {Balents}},\ }\href
  {\doibase 10.1038/nphys2465} {\bibfield  {journal} {\bibinfo  {journal} {Nat.
  Phys.}\ }\textbf {\bibinfo {volume} {8}},\ \bibinfo {pages} {902} (\bibinfo
  {year} {2012})}\BibitemShut {NoStop}%
\bibitem [{\citenamefont {He}\ \emph {et~al.}(2017{\natexlab{a}})\citenamefont
  {He}, \citenamefont {Zaletel}, \citenamefont {Oshikawa},\ and\ \citenamefont
  {Pollmann}}]{He2017}%
  \BibitemOpen
  \bibfield  {author} {\bibinfo {author} {\bibfnamefont {Y.-C.}\ \bibnamefont
  {He}}, \bibinfo {author} {\bibfnamefont {M.~P.}\ \bibnamefont {Zaletel}},
  \bibinfo {author} {\bibfnamefont {M.}~\bibnamefont {Oshikawa}}, \ and\
  \bibinfo {author} {\bibfnamefont {F.}~\bibnamefont {Pollmann}},\ }\href
  {\doibase 10.1103/PhysRevX.7.031020} {\bibfield  {journal} {\bibinfo
  {journal} {Phys. Rev. X}\ }\textbf {\bibinfo {volume} {7}},\ \bibinfo {pages}
  {031020} (\bibinfo {year} {2017}{\natexlab{a}})}\BibitemShut {NoStop}%
\bibitem [{\citenamefont {Norman}(2016)}]{Norman2016}%
  \BibitemOpen
  \bibfield  {author} {\bibinfo {author} {\bibfnamefont {M.~R.}\ \bibnamefont
  {Norman}},\ }\href {\doibase 10.1103/RevModPhys.88.041002} {\bibfield
  {journal} {\bibinfo  {journal} {Rev. Mod. Phys.}\ }\textbf {\bibinfo {volume}
  {88}},\ \bibinfo {pages} {041002} (\bibinfo {year} {2016})}\BibitemShut
  {NoStop}%
\bibitem [{\citenamefont {Schnyder}\ \emph {et~al.}(2008)\citenamefont
  {Schnyder}, \citenamefont {Starykh},\ and\ \citenamefont
  {Balents}}]{Schnyder2008}%
  \BibitemOpen
  \bibfield  {author} {\bibinfo {author} {\bibfnamefont {A.~P.}\ \bibnamefont
  {Schnyder}}, \bibinfo {author} {\bibfnamefont {O.~A.}\ \bibnamefont
  {Starykh}}, \ and\ \bibinfo {author} {\bibfnamefont {L.}~\bibnamefont
  {Balents}},\ }\href {\doibase 10.1103/PhysRevB.78.174420} {\bibfield
  {journal} {\bibinfo  {journal} {Phys. Rev. B}\ }\textbf {\bibinfo {volume}
  {78}},\ \bibinfo {pages} {174420} (\bibinfo {year} {2008})}\BibitemShut
  {NoStop}%
\bibitem [{\citenamefont {Mila}(1998)}]{Mila98}%
  \BibitemOpen
  \bibfield  {author} {\bibinfo {author} {\bibfnamefont {F.}~\bibnamefont
  {Mila}},\ }\href {\doibase 10.1103/PhysRevLett.81.2356} {\bibfield  {journal}
  {\bibinfo  {journal} {Phys. Rev. Lett.}\ }\textbf {\bibinfo {volume} {81}},\
  \bibinfo {pages} {2356} (\bibinfo {year} {1998})}\BibitemShut {NoStop}%
\bibitem [{\citenamefont {Aidoudi}\ \emph {et~al.}(2011)\citenamefont
  {Aidoudi}, \citenamefont {Aldous}, \citenamefont {Goff}, \citenamefont
  {Slawin}, \citenamefont {Attfield}, \citenamefont {Morris},\ and\
  \citenamefont {Lightfoot}}]{Aidoudi2011}%
  \BibitemOpen
  \bibfield  {author} {\bibinfo {author} {\bibfnamefont {F.~H.}\ \bibnamefont
  {Aidoudi}}, \bibinfo {author} {\bibfnamefont {D.~W.}\ \bibnamefont {Aldous}},
  \bibinfo {author} {\bibfnamefont {R.~J.}\ \bibnamefont {Goff}}, \bibinfo
  {author} {\bibfnamefont {A.~M.~Z.}\ \bibnamefont {Slawin}}, \bibinfo {author}
  {\bibfnamefont {J.~P.}\ \bibnamefont {Attfield}}, \bibinfo {author}
  {\bibfnamefont {R.~E.}\ \bibnamefont {Morris}}, \ and\ \bibinfo {author}
  {\bibfnamefont {P.}~\bibnamefont {Lightfoot}},\ }\href {\doibase
  10.1038/nchem.1129} {\bibfield  {journal} {\bibinfo  {journal} {Nat. Chem.}\
  }\textbf {\bibinfo {volume} {3}},\ \bibinfo {pages} {801} (\bibinfo {year}
  {2011})}\BibitemShut {NoStop}%
\bibitem [{\citenamefont {Honecker}\ \emph {et~al.}(2011)\citenamefont
  {Honecker}, \citenamefont {Cabra}, \citenamefont {Everts}, \citenamefont
  {Pujol},\ and\ \citenamefont {Stauffer}}]{Honecker11}%
  \BibitemOpen
  \bibfield  {author} {\bibinfo {author} {\bibfnamefont {A.}~\bibnamefont
  {Honecker}}, \bibinfo {author} {\bibfnamefont {D.~C.}\ \bibnamefont {Cabra}},
  \bibinfo {author} {\bibfnamefont {H.-U.}\ \bibnamefont {Everts}}, \bibinfo
  {author} {\bibfnamefont {P.}~\bibnamefont {Pujol}}, \ and\ \bibinfo {author}
  {\bibfnamefont {F.}~\bibnamefont {Stauffer}},\ }\href {\doibase
  10.1103/PhysRevB.84.224410} {\bibfield  {journal} {\bibinfo  {journal} {Phys.
  Rev. B}\ }\textbf {\bibinfo {volume} {84}},\ \bibinfo {pages} {224410}
  (\bibinfo {year} {2011})}\BibitemShut {NoStop}%
\bibitem [{\citenamefont {Repellin}\ \emph {et~al.}(2017)\citenamefont
  {Repellin}, \citenamefont {He},\ and\ \citenamefont
  {Pollmann}}]{Repellin2017}%
  \BibitemOpen
  \bibfield  {author} {\bibinfo {author} {\bibfnamefont {C.}~\bibnamefont
  {Repellin}}, \bibinfo {author} {\bibfnamefont {Y.-C.}\ \bibnamefont {He}}, \
  and\ \bibinfo {author} {\bibfnamefont {F.}~\bibnamefont {Pollmann}},\ }\href
  {\doibase 10.1103/PhysRevB.96.205124} {\bibfield  {journal} {\bibinfo
  {journal} {Phys. Rev. B}\ }\textbf {\bibinfo {volume} {96}},\ \bibinfo
  {pages} {205124} (\bibinfo {year} {2017})}\BibitemShut {NoStop}%
\bibitem [{\citenamefont {Jahromi}\ \emph {et~al.}(2020)\citenamefont
  {Jahromi}, \citenamefont {Orús}, \citenamefont {Poilblanc},\ and\
  \citenamefont {Mila}}]{Jahromi2020}%
  \BibitemOpen
  \bibfield  {author} {\bibinfo {author} {\bibfnamefont {S.~S.}\ \bibnamefont
  {Jahromi}}, \bibinfo {author} {\bibfnamefont {R.}~\bibnamefont {Orús}},
  \bibinfo {author} {\bibfnamefont {D.}~\bibnamefont {Poilblanc}}, \ and\
  \bibinfo {author} {\bibfnamefont {F.}~\bibnamefont {Mila}},\ }\href {\doibase
  10.21468/SciPostPhys.9.6.092} {\bibfield  {journal} {\bibinfo  {journal}
  {SciPost Phys.}\ }\textbf {\bibinfo {volume} {9}},\ \bibinfo {pages} {92}
  (\bibinfo {year} {2020})}\BibitemShut {NoStop}%
\bibitem [{\citenamefont {Clark}\ \emph {et~al.}(2013)\citenamefont {Clark},
  \citenamefont {Orain}, \citenamefont {Bert}, \citenamefont {De~Vries},
  \citenamefont {Aidoudi}, \citenamefont {Morris}, \citenamefont {Lightfoot},
  \citenamefont {Lord}, \citenamefont {Telling}, \citenamefont {Bonville},
  \citenamefont {Attfield}, \citenamefont {Mendels},\ and\ \citenamefont
  {Harrison}}]{Clark2013}%
  \BibitemOpen
  \bibfield  {author} {\bibinfo {author} {\bibfnamefont {L.}~\bibnamefont
  {Clark}}, \bibinfo {author} {\bibfnamefont {J.~C.}\ \bibnamefont {Orain}},
  \bibinfo {author} {\bibfnamefont {F.}~\bibnamefont {Bert}}, \bibinfo {author}
  {\bibfnamefont {M.~A.}\ \bibnamefont {De~Vries}}, \bibinfo {author}
  {\bibfnamefont {F.~H.}\ \bibnamefont {Aidoudi}}, \bibinfo {author}
  {\bibfnamefont {R.~E.}\ \bibnamefont {Morris}}, \bibinfo {author}
  {\bibfnamefont {P.}~\bibnamefont {Lightfoot}}, \bibinfo {author}
  {\bibfnamefont {J.~S.}\ \bibnamefont {Lord}}, \bibinfo {author}
  {\bibfnamefont {M.~T.~F.}\ \bibnamefont {Telling}}, \bibinfo {author}
  {\bibfnamefont {P.}~\bibnamefont {Bonville}}, \bibinfo {author}
  {\bibfnamefont {J.~P.}\ \bibnamefont {Attfield}}, \bibinfo {author}
  {\bibfnamefont {P.}~\bibnamefont {Mendels}}, \ and\ \bibinfo {author}
  {\bibfnamefont {A.}~\bibnamefont {Harrison}},\ }\href {\doibase
  10.1103/PhysRevLett.110.207208} {\bibfield  {journal} {\bibinfo  {journal}
  {Phys. Rev. Lett.}\ }\textbf {\bibinfo {volume} {110}},\ \bibinfo {pages}
  {207208} (\bibinfo {year} {2013})}\BibitemShut {NoStop}%
\bibitem [{\citenamefont {Orain}\ \emph {et~al.}(2017)\citenamefont {Orain},
  \citenamefont {Bernu}, \citenamefont {Mendels}, \citenamefont {Clark},
  \citenamefont {Aidoudi}, \citenamefont {Lightfoot}, \citenamefont {Morris},\
  and\ \citenamefont {Bert}}]{Orain2017}%
  \BibitemOpen
  \bibfield  {author} {\bibinfo {author} {\bibfnamefont {J.-C.}\ \bibnamefont
  {Orain}}, \bibinfo {author} {\bibfnamefont {B.}~\bibnamefont {Bernu}},
  \bibinfo {author} {\bibfnamefont {P.}~\bibnamefont {Mendels}}, \bibinfo
  {author} {\bibfnamefont {L.}~\bibnamefont {Clark}}, \bibinfo {author}
  {\bibfnamefont {F.~H.}\ \bibnamefont {Aidoudi}}, \bibinfo {author}
  {\bibfnamefont {P.}~\bibnamefont {Lightfoot}}, \bibinfo {author}
  {\bibfnamefont {R.~E.}\ \bibnamefont {Morris}}, \ and\ \bibinfo {author}
  {\bibfnamefont {F.}~\bibnamefont {Bert}},\ }\href {\doibase
  10.1103/PhysRevLett.118.237203} {\bibfield  {journal} {\bibinfo  {journal}
  {Phys. Rev. Lett.}\ }\textbf {\bibinfo {volume} {118}},\ \bibinfo {pages}
  {237203} (\bibinfo {year} {2017})}\BibitemShut {NoStop}%
\bibitem [{\citenamefont {Sugiura}\ and\ \citenamefont
  {Shimizu}(2013)}]{Sugiura13}%
  \BibitemOpen
  \bibfield  {author} {\bibinfo {author} {\bibfnamefont {S.}~\bibnamefont
  {Sugiura}}\ and\ \bibinfo {author} {\bibfnamefont {A.}~\bibnamefont
  {Shimizu}},\ }\href {\doibase 10.1103/PhysRevLett.111.010401} {\bibfield
  {journal} {\bibinfo  {journal} {Phys. Rev. Lett.}\ }\textbf {\bibinfo
  {volume} {111}},\ \bibinfo {pages} {010401} (\bibinfo {year}
  {2013})}\BibitemShut {NoStop}%
\bibitem [{\citenamefont {L\"auchli}\ \emph {et~al.}(2019)\citenamefont
  {L\"auchli}, \citenamefont {Sudan},\ and\ \citenamefont
  {Moessner}}]{Laeuchli19}%
  \BibitemOpen
  \bibfield  {author} {\bibinfo {author} {\bibfnamefont {A.~M.}\ \bibnamefont
  {L\"auchli}}, \bibinfo {author} {\bibfnamefont {J.}~\bibnamefont {Sudan}}, \
  and\ \bibinfo {author} {\bibfnamefont {R.}~\bibnamefont {Moessner}},\ }\href
  {\doibase 10.1103/PhysRevB.100.155142} {\bibfield  {journal} {\bibinfo
  {journal} {Phys. Rev. B}\ }\textbf {\bibinfo {volume} {100}},\ \bibinfo
  {pages} {155142} (\bibinfo {year} {2019})}\BibitemShut {NoStop}%
\bibitem [{\citenamefont {Schnack}\ \emph {et~al.}(2018)\citenamefont
  {Schnack}, \citenamefont {Schulenburg},\ and\ \citenamefont
  {Richter}}]{Schnack18}%
  \BibitemOpen
  \bibfield  {author} {\bibinfo {author} {\bibfnamefont {J.}~\bibnamefont
  {Schnack}}, \bibinfo {author} {\bibfnamefont {J.}~\bibnamefont
  {Schulenburg}}, \ and\ \bibinfo {author} {\bibfnamefont {J.}~\bibnamefont
  {Richter}},\ }\href {\doibase 10.1103/PhysRevB.98.094423} {\bibfield
  {journal} {\bibinfo  {journal} {Phys. Rev. B}\ }\textbf {\bibinfo {volume}
  {98}},\ \bibinfo {pages} {094423} (\bibinfo {year} {2018})}\BibitemShut
  {NoStop}%
\bibitem [{\citenamefont {Yan}\ \emph {et~al.}(2011{\natexlab{b}})\citenamefont
  {Yan}, \citenamefont {Huse},\ and\ \citenamefont {White}}]{Yan11}%
  \BibitemOpen
  \bibfield  {author} {\bibinfo {author} {\bibfnamefont {S.}~\bibnamefont
  {Yan}}, \bibinfo {author} {\bibfnamefont {D.~A.}\ \bibnamefont {Huse}}, \
  and\ \bibinfo {author} {\bibfnamefont {S.~R.}\ \bibnamefont {White}},\ }\href
  {\doibase 10.1126/science.1201080} {\bibfield  {journal} {\bibinfo  {journal}
  {Science}\ }\textbf {\bibinfo {volume} {332}},\ \bibinfo {pages} {1173}
  (\bibinfo {year} {2011}{\natexlab{b}})}\BibitemShut {NoStop}%
\bibitem [{\citenamefont {He}\ \emph {et~al.}(2017{\natexlab{b}})\citenamefont
  {He}, \citenamefont {Zaletel}, \citenamefont {Oshikawa},\ and\ \citenamefont
  {Pollmann}}]{He17}%
  \BibitemOpen
  \bibfield  {author} {\bibinfo {author} {\bibfnamefont {Y.-C.}\ \bibnamefont
  {He}}, \bibinfo {author} {\bibfnamefont {M.~P.}\ \bibnamefont {Zaletel}},
  \bibinfo {author} {\bibfnamefont {M.}~\bibnamefont {Oshikawa}}, \ and\
  \bibinfo {author} {\bibfnamefont {F.}~\bibnamefont {Pollmann}},\ }\href
  {\doibase 10.1103/PhysRevX.7.031020} {\bibfield  {journal} {\bibinfo
  {journal} {Phys. Rev. X}\ }\textbf {\bibinfo {volume} {7}},\ \bibinfo {pages}
  {031020} (\bibinfo {year} {2017}{\natexlab{b}})}\BibitemShut {NoStop}%
\bibitem [{\citenamefont {Liao}\ \emph {et~al.}(2017)\citenamefont {Liao},
  \citenamefont {Xie}, \citenamefont {Chen}, \citenamefont {Liu}, \citenamefont
  {Xie}, \citenamefont {Huang}, \citenamefont {Normand},\ and\ \citenamefont
  {Xiang}}]{Liao17}%
  \BibitemOpen
  \bibfield  {author} {\bibinfo {author} {\bibfnamefont {H.~J.}\ \bibnamefont
  {Liao}}, \bibinfo {author} {\bibfnamefont {Z.~Y.}\ \bibnamefont {Xie}},
  \bibinfo {author} {\bibfnamefont {J.}~\bibnamefont {Chen}}, \bibinfo {author}
  {\bibfnamefont {Z.~Y.}\ \bibnamefont {Liu}}, \bibinfo {author} {\bibfnamefont
  {H.~D.}\ \bibnamefont {Xie}}, \bibinfo {author} {\bibfnamefont {R.~Z.}\
  \bibnamefont {Huang}}, \bibinfo {author} {\bibfnamefont {B.}~\bibnamefont
  {Normand}}, \ and\ \bibinfo {author} {\bibfnamefont {T.}~\bibnamefont
  {Xiang}},\ }\href {\doibase 10.1103/PhysRevLett.118.137202} {\bibfield
  {journal} {\bibinfo  {journal} {Phys. Rev. Lett.}\ }\textbf {\bibinfo
  {volume} {118}},\ \bibinfo {pages} {137202} (\bibinfo {year}
  {2017})}\BibitemShut {NoStop}%
\bibitem [{\citenamefont {Chen}\ \emph {et~al.}(2018)\citenamefont {Chen},
  \citenamefont {Ran}, \citenamefont {Liu}, \citenamefont {Peng}, \citenamefont
  {Huang},\ and\ \citenamefont {Su}}]{Chen18}%
  \BibitemOpen
  \bibfield  {author} {\bibinfo {author} {\bibfnamefont {X.}~\bibnamefont
  {Chen}}, \bibinfo {author} {\bibfnamefont {S.-J.}\ \bibnamefont {Ran}},
  \bibinfo {author} {\bibfnamefont {T.}~\bibnamefont {Liu}}, \bibinfo {author}
  {\bibfnamefont {C.}~\bibnamefont {Peng}}, \bibinfo {author} {\bibfnamefont
  {Y.-Z.}\ \bibnamefont {Huang}}, \ and\ \bibinfo {author} {\bibfnamefont
  {G.}~\bibnamefont {Su}},\ }\href {\doibase
  https://doi.org/10.1016/j.scib.2018.11.007} {\bibfield  {journal} {\bibinfo
  {journal} {Science Bulletin}\ }\textbf {\bibinfo {volume} {63}},\ \bibinfo
  {pages} {1545} (\bibinfo {year} {2018})}\BibitemShut {NoStop}%
\bibitem [{\citenamefont {Sandvik}\ and\ \citenamefont
  {Kurkij\"arvi}(1991)}]{Sandvik1991}%
  \BibitemOpen
  \bibfield  {author} {\bibinfo {author} {\bibfnamefont {A.~W.}\ \bibnamefont
  {Sandvik}}\ and\ \bibinfo {author} {\bibfnamefont {J.}~\bibnamefont
  {Kurkij\"arvi}},\ }\href {\doibase 10.1103/PhysRevB.43.5950} {\bibfield
  {journal} {\bibinfo  {journal} {Phys. Rev. B}\ }\textbf {\bibinfo {volume}
  {43}},\ \bibinfo {pages} {5950} (\bibinfo {year} {1991})}\BibitemShut
  {NoStop}%
\bibitem [{\citenamefont {Sandvik}(1992)}]{Sandvik1992}%
  \BibitemOpen
  \bibfield  {author} {\bibinfo {author} {\bibfnamefont {A.~W.}\ \bibnamefont
  {Sandvik}},\ }\href {https://doi.org/10.1088/0305-4470/25/13/017} {\bibfield
  {journal} {\bibinfo  {journal} {J. Phys. A.}\ }\textbf {\bibinfo {volume}
  {25}},\ \bibinfo {pages} {3667} (\bibinfo {year} {1992})}\BibitemShut
  {NoStop}%
\bibitem [{\citenamefont {Sandvik}(1999)}]{Sandvik1999}%
  \BibitemOpen
  \bibfield  {author} {\bibinfo {author} {\bibfnamefont {A.~W.}\ \bibnamefont
  {Sandvik}},\ }\href {\doibase 10.1103/PhysRevB.59.R14157} {\bibfield
  {journal} {\bibinfo  {journal} {Phys. Rev. B}\ }\textbf {\bibinfo {volume}
  {59}},\ \bibinfo {pages} {R14157} (\bibinfo {year} {1999})}\BibitemShut
  {NoStop}%
\bibitem [{\citenamefont {Sylju\aa{}sen}\ and\ \citenamefont
  {Sandvik}(2002)}]{Syljuasen2002}%
  \BibitemOpen
  \bibfield  {author} {\bibinfo {author} {\bibfnamefont {O.~F.}\ \bibnamefont
  {Sylju\aa{}sen}}\ and\ \bibinfo {author} {\bibfnamefont {A.~W.}\ \bibnamefont
  {Sandvik}},\ }\href {\doibase 10.1103/PhysRevE.66.046701} {\bibfield
  {journal} {\bibinfo  {journal} {Phys. Rev. E}\ }\textbf {\bibinfo {volume}
  {66}},\ \bibinfo {pages} {046701} (\bibinfo {year} {2002})}\BibitemShut
  {NoStop}%
\bibitem [{\citenamefont {Alet}\ \emph {et~al.}(2005)\citenamefont {Alet},
  \citenamefont {Wessel},\ and\ \citenamefont {Troyer}}]{Alet2005}%
  \BibitemOpen
  \bibfield  {author} {\bibinfo {author} {\bibfnamefont {F.}~\bibnamefont
  {Alet}}, \bibinfo {author} {\bibfnamefont {S.}~\bibnamefont {Wessel}}, \ and\
  \bibinfo {author} {\bibfnamefont {M.}~\bibnamefont {Troyer}},\ }\href
  {\doibase 10.1103/PhysRevE.71.036706} {\bibfield  {journal} {\bibinfo
  {journal} {Phys. Rev. E}\ }\textbf {\bibinfo {volume} {71}},\ \bibinfo
  {pages} {036706} (\bibinfo {year} {2005})}\BibitemShut {NoStop}%
\bibitem [{\citenamefont {Henelius}\ and\ \citenamefont
  {Sandvik}(2000)}]{Henelius00}%
  \BibitemOpen
  \bibfield  {author} {\bibinfo {author} {\bibfnamefont {P.}~\bibnamefont
  {Henelius}}\ and\ \bibinfo {author} {\bibfnamefont {A.~W.}\ \bibnamefont
  {Sandvik}},\ }\href {\doibase 10.1103/PhysRevB.62.1102} {\bibfield  {journal}
  {\bibinfo  {journal} {Phys. Rev. B}\ }\textbf {\bibinfo {volume} {62}},\
  \bibinfo {pages} {1102} (\bibinfo {year} {2000})}\BibitemShut {NoStop}%
\bibitem [{\citenamefont {Troyer}\ and\ \citenamefont
  {Wiese}(2005)}]{Troyer2005}%
  \BibitemOpen
  \bibfield  {author} {\bibinfo {author} {\bibfnamefont {M.}~\bibnamefont
  {Troyer}}\ and\ \bibinfo {author} {\bibfnamefont {U.-J.}\ \bibnamefont
  {Wiese}},\ }\href {\doibase 10.1103/PhysRevLett.94.170201} {\bibfield
  {journal} {\bibinfo  {journal} {Phys. Rev. Lett.}\ }\textbf {\bibinfo
  {volume} {94}},\ \bibinfo {pages} {170201} (\bibinfo {year}
  {2005})}\BibitemShut {NoStop}%
\bibitem [{\citenamefont {Nakamura}(1998)}]{Nakamura1998}%
  \BibitemOpen
  \bibfield  {author} {\bibinfo {author} {\bibfnamefont {T.}~\bibnamefont
  {Nakamura}},\ }\href {\doibase 10.1103/PhysRevB.57.R3197} {\bibfield
  {journal} {\bibinfo  {journal} {Phys. Rev. B}\ }\textbf {\bibinfo {volume}
  {57}},\ \bibinfo {pages} {R3197} (\bibinfo {year} {1998})}\BibitemShut
  {NoStop}%
\bibitem [{\citenamefont {Alet}\ \emph {et~al.}(2016)\citenamefont {Alet},
  \citenamefont {Damle},\ and\ \citenamefont {Pujari}}]{Alet2016}%
  \BibitemOpen
  \bibfield  {author} {\bibinfo {author} {\bibfnamefont {F.}~\bibnamefont
  {Alet}}, \bibinfo {author} {\bibfnamefont {K.}~\bibnamefont {Damle}}, \ and\
  \bibinfo {author} {\bibfnamefont {S.}~\bibnamefont {Pujari}},\ }\href
  {\doibase 10.1103/PhysRevLett.117.197203} {\bibfield  {journal} {\bibinfo
  {journal} {Phys. Rev. Lett.}\ }\textbf {\bibinfo {volume} {117}},\ \bibinfo
  {pages} {197203} (\bibinfo {year} {2016})}\BibitemShut {NoStop}%
\bibitem [{\citenamefont {Honecker}\ \emph {et~al.}(2016)\citenamefont
  {Honecker}, \citenamefont {Wessel}, \citenamefont {Kerkdyk}, \citenamefont
  {Pruschke}, \citenamefont {Mila},\ and\ \citenamefont
  {Normand}}]{Honecker2016}%
  \BibitemOpen
  \bibfield  {author} {\bibinfo {author} {\bibfnamefont {A.}~\bibnamefont
  {Honecker}}, \bibinfo {author} {\bibfnamefont {S.}~\bibnamefont {Wessel}},
  \bibinfo {author} {\bibfnamefont {R.}~\bibnamefont {Kerkdyk}}, \bibinfo
  {author} {\bibfnamefont {T.}~\bibnamefont {Pruschke}}, \bibinfo {author}
  {\bibfnamefont {F.}~\bibnamefont {Mila}}, \ and\ \bibinfo {author}
  {\bibfnamefont {B.}~\bibnamefont {Normand}},\ }\href {\doibase
  10.1103/PhysRevB.93.054408} {\bibfield  {journal} {\bibinfo  {journal} {Phys.
  Rev. B}\ }\textbf {\bibinfo {volume} {93}},\ \bibinfo {pages} {054408}
  (\bibinfo {year} {2016})}\BibitemShut {NoStop}%
\bibitem [{\citenamefont {Ng}\ and\ \citenamefont {Yang}(2017)}]{Ng2017}%
  \BibitemOpen
  \bibfield  {author} {\bibinfo {author} {\bibfnamefont {K.-K.}\ \bibnamefont
  {Ng}}\ and\ \bibinfo {author} {\bibfnamefont {M.-F.}\ \bibnamefont {Yang}},\
  }\href {\doibase 10.1103/PhysRevB.95.064431} {\bibfield  {journal} {\bibinfo
  {journal} {Phys. Rev. B}\ }\textbf {\bibinfo {volume} {95}},\ \bibinfo
  {pages} {064431} (\bibinfo {year} {2017})}\BibitemShut {NoStop}%
\bibitem [{\citenamefont {Stapmanns}\ \emph {et~al.}(2018)\citenamefont
  {Stapmanns}, \citenamefont {Corboz}, \citenamefont {Mila}, \citenamefont
  {Honecker}, \citenamefont {Normand},\ and\ \citenamefont
  {Wessel}}]{Stapmanns2018}%
  \BibitemOpen
  \bibfield  {author} {\bibinfo {author} {\bibfnamefont {J.}~\bibnamefont
  {Stapmanns}}, \bibinfo {author} {\bibfnamefont {P.}~\bibnamefont {Corboz}},
  \bibinfo {author} {\bibfnamefont {F.}~\bibnamefont {Mila}}, \bibinfo {author}
  {\bibfnamefont {A.}~\bibnamefont {Honecker}}, \bibinfo {author}
  {\bibfnamefont {B.}~\bibnamefont {Normand}}, \ and\ \bibinfo {author}
  {\bibfnamefont {S.}~\bibnamefont {Wessel}},\ }\href {\doibase
  10.1103/PhysRevLett.121.127201} {\bibfield  {journal} {\bibinfo  {journal}
  {Phys. Rev. Lett.}\ }\textbf {\bibinfo {volume} {121}},\ \bibinfo {pages}
  {127201} (\bibinfo {year} {2018})}\BibitemShut {NoStop}%
\bibitem [{\citenamefont {Weber}\ \emph {et~al.}(2022)\citenamefont {Weber},
  \citenamefont {Honecker}, \citenamefont {Normand}, \citenamefont {Corboz},
  \citenamefont {Mila},\ and\ \citenamefont {Wessel}}]{Weber2021}%
  \BibitemOpen
  \bibfield  {author} {\bibinfo {author} {\bibfnamefont {L.}~\bibnamefont
  {Weber}}, \bibinfo {author} {\bibfnamefont {A.}~\bibnamefont {Honecker}},
  \bibinfo {author} {\bibfnamefont {B.}~\bibnamefont {Normand}}, \bibinfo
  {author} {\bibfnamefont {P.}~\bibnamefont {Corboz}}, \bibinfo {author}
  {\bibfnamefont {F.}~\bibnamefont {Mila}}, \ and\ \bibinfo {author}
  {\bibfnamefont {S.}~\bibnamefont {Wessel}},\ }\href {\doibase
  10.21468/SciPostPhys.12.2.054} {\bibfield  {journal} {\bibinfo  {journal}
  {SciPost Phys.}\ }\textbf {\bibinfo {volume} {12}},\ \bibinfo {pages} {54}
  (\bibinfo {year} {2022})}\BibitemShut {NoStop}%
\bibitem [{\citenamefont {Wessel}\ and\ \citenamefont
  {Haas}(2001)}]{Wessel2001}%
  \BibitemOpen
  \bibfield  {author} {\bibinfo {author} {\bibfnamefont {S.}~\bibnamefont
  {Wessel}}\ and\ \bibinfo {author} {\bibfnamefont {S.}~\bibnamefont {Haas}},\
  }\href {\doibase 10.1103/PhysRevB.63.140403} {\bibfield  {journal} {\bibinfo
  {journal} {Phys. Rev. B}\ }\textbf {\bibinfo {volume} {63}},\ \bibinfo
  {pages} {140403} (\bibinfo {year} {2001})}\BibitemShut {NoStop}%
\bibitem [{\citenamefont {Wang}(2001)}]{Wang2001}%
  \BibitemOpen
  \bibfield  {author} {\bibinfo {author} {\bibfnamefont {H.-T.}\ \bibnamefont
  {Wang}},\ }\href {\doibase 10.1103/PhysRevB.65.024426} {\bibfield  {journal}
  {\bibinfo  {journal} {Phys. Rev. B}\ }\textbf {\bibinfo {volume} {65}},\
  \bibinfo {pages} {024426} (\bibinfo {year} {2001})}\BibitemShut {NoStop}%
\bibitem [{Note1()}]{Note1}%
  \BibitemOpen
  \bibinfo {note} {Even with a complex basis, one can, in principle, retain
  real-valued weights within the SSE and related algorithms due to pairs of
  configurations with cancelling imaginary parts \cite {Hen2021}. In this work,
  however, we concentrate on real-valued bases.}\BibitemShut {Stop}%
\bibitem [{\citenamefont {Subrahmanyam}(1995)}]{Subrahmanyam95}%
  \BibitemOpen
  \bibfield  {author} {\bibinfo {author} {\bibfnamefont {V.}~\bibnamefont
  {Subrahmanyam}},\ }\href {\doibase 10.1103/PhysRevB.52.1133} {\bibfield
  {journal} {\bibinfo  {journal} {Phys. Rev. B}\ }\textbf {\bibinfo {volume}
  {52}},\ \bibinfo {pages} {1133} (\bibinfo {year} {1995})}\BibitemShut
  {NoStop}%
\bibitem [{Note2()}]{Note2}%
  \BibitemOpen
  \bibinfo {note} {Because our spin numbering convention differs from
  Ref.~\cite {Wang2001}, we pick up an extra phase $\omega $ here, simplifying
  the Hamiltonian slightly.}\BibitemShut {Stop}%
\bibitem [{\citenamefont {Misguich}\ and\ \citenamefont
  {Bernu}(2005)}]{misguich05}%
  \BibitemOpen
  \bibfield  {author} {\bibinfo {author} {\bibfnamefont {G.}~\bibnamefont
  {Misguich}}\ and\ \bibinfo {author} {\bibfnamefont {B.}~\bibnamefont
  {Bernu}},\ }\href@noop {} {\bibfield  {journal} {\bibinfo  {journal} {Phys.
  Rev. B}\ }\textbf {\bibinfo {volume} {71}},\ \bibinfo {pages} {014417}
  (\bibinfo {year} {2005})}\BibitemShut {NoStop}%
\bibitem [{\citenamefont {Shimokawa}\ and\ \citenamefont
  {Kawamura}(2016)}]{Shimokawa16}%
  \BibitemOpen
  \bibfield  {author} {\bibinfo {author} {\bibfnamefont {T.}~\bibnamefont
  {Shimokawa}}\ and\ \bibinfo {author} {\bibfnamefont {H.}~\bibnamefont
  {Kawamura}},\ }\href {\doibase 10.7566/JPSJ.85.113702} {\bibfield  {journal}
  {\bibinfo  {journal} {J. Phys. Soc. Jpn.}\ }\textbf {\bibinfo {volume}
  {85}},\ \bibinfo {pages} {113702} (\bibinfo {year} {2016})}\BibitemShut
  {NoStop}%
\bibitem [{\citenamefont {Dorneich}\ and\ \citenamefont
  {Troyer}(2001)}]{Dorneich2001}%
  \BibitemOpen
  \bibfield  {author} {\bibinfo {author} {\bibfnamefont {A.}~\bibnamefont
  {Dorneich}}\ and\ \bibinfo {author} {\bibfnamefont {M.}~\bibnamefont
  {Troyer}},\ }\href {\doibase 10.1103/PhysRevE.64.066701} {\bibfield
  {journal} {\bibinfo  {journal} {Phys. Rev. E}\ }\textbf {\bibinfo {volume}
  {64}},\ \bibinfo {pages} {066701} (\bibinfo {year} {2001})}\BibitemShut
  {NoStop}%
\bibitem [{\citenamefont {Hen}(2021)}]{Hen2021}%
  \BibitemOpen
  \bibfield  {author} {\bibinfo {author} {\bibfnamefont {I.}~\bibnamefont
  {Hen}},\ }\href {\doibase 10.1103/PhysRevResearch.3.023080} {\bibfield
  {journal} {\bibinfo  {journal} {Phys. Rev. Research}\ }\textbf {\bibinfo
  {volume} {3}},\ \bibinfo {pages} {023080} (\bibinfo {year}
  {2021})}\BibitemShut {NoStop}%
\end{thebibliography}%

\end{document}